# Review:

# Adaptive Radiation Therapy for Head and Neck Cancer

Lucas McCullum[1,2] (lbmccullum@mdanderson.org, 0000-0001-9788-7987)
Sonali J. Joshi[3] (sonali.joshi@md.cusm.edu, 0009-0003-8230-172X)
Brandon M. Godinich[4] (brandongodinich@gmail.com, 0000-0002-4525-1693)
Parshawn Gerafian[5] (parshangeraf81@berkeley.edu, 0009-0003-9975-9139)
Rishabh Gaur[6] (rgfd8@umkc.edu, 0009-0006-3200-4585)
Qusai Alakayleh[2,7] (qmalakayleh@mdanderson.org, 0009-0005-0050-6567)
Ergys Subashi[8] (edsubashi@mdanderson.org, 0000-0001-5168-6928)
Renjie He[2] (rhe1@mdanderson.org, 0000-0001-9166-6286)
Samuel L. Mulder[1,2] (smulder@mdanderson.org, 0000-0001-5185-4805)
Zaphanlene Kaffey[1,2] (zkaffey@mdanderson.org, 0009-0008-7999-5245)
Grace Murley[1,2] (gisakson@mdanderson.org, 0000-0003-0736-714X)
Natalie A. West[1,2] (nawest@mdanderson.org, 0009-0007-0211-8845)
Saleh Ramezani[2] (sramezani@mdanderson.org, 0000-0002-7872-9454)
Cem Dede[1,2] (cdede@mdanderson.org, 0000-0002-0543-9325)
Laia Humbert-Vidan[2] (lhumbert@mdanderson.org, 0000-0002-8005-6770)
Clifton D. Fuller[1,2] (cdfuller@mdanderson.org, 0000-0002-5264-3994)

[1]UT MD Anderson Cancer Center UTHealth Houston Graduate School of Biomedical Sciences, Houston, USA
[2]Department of Radiation Oncology, The University of Texas MD Anderson Cancer Center, Houston, TX, USA
[3]California University of Science and Medicine, Colton, CA, USA
[4]Paul L. Foster School of Medicine, Texas Tech University Health Sciences Center El Paso, El Paso, TX, USA
[5]University of California at Berkeley, Berkeley, CA, USA
[6]University of Missouri-Kansas City School of Medicine, Kansas City, MO, USA
[7]Department of Electrical and Computer Engineering, Lamar University, Beaumont, TX, USA
[8]Department of Radiation Physics, The University of Texas MD Anderson Cancer Center, Houston, TX, USA

**Funding Statement**

LM is supported by a National Institutes of Health (NIH) Diversity Supplement (R01CA257814-02S2). SM and GM were supported by a training fellowship from UTHealth Houston Center for Clinical and Translational Sciences TL1 Program (TL1 TR003169). ZK was supported by a pre-doctoral fellowship from the Cancer Prevention Research Institute of Texas grant #RP210042. NAW is supported by a training fellowship from UTHealth Houston Center for Clinical and Translational Sciences T32 Program (Grant No. T32 TR004905) and a NIH National Institute of Dental and Craniofacial Research (NIDCR) Academic Industrial Partnership Grant (R01DE028290). CDF has received unrelated funding and salary support from: NIH National Institute of Dental and Craniofacial Research (NIDCR) Academic Industrial Partnership (R01DE028290), the Administrative Supplement to Support Collaborations to Improve AIML-Readiness of NIH-Supported Data (R01DE028290-04S2); NIDCR Establishing Outcome Measures for Clinical Studies of Oral and Craniofacial Diseases and Conditions award (R01DE025248); NSF/NIH Interagency Smart and Connected Health (SCH) Program (R01CA257814); NIH National Institute of Biomedical Imaging and Bioengineering (NIBIB) Research Education Programs for Residents and Clinical Fellows Grant (R25EB025787); NIH NIDCR Exploratory/Developmental Research Grant Program (R21DE031082); NIH/NCI Cancer Center Support Grant (CCSG) Pilot Research Program Award from the UT MD Anderson CCSG Radiation Oncology and Cancer Imaging Program (P30CA016672); Patient-Centered Outcomes Research Institute (PCS-1609-36195) sub-award from Princess Margaret Hospital; National Science Foundation (NSF) Division of Civil, Mechanical, and Manufacturing Innovation (CMMI) grant (NSF 1933369). CDF receives grant and infrastructure support from MD Anderson Cancer Center via: the Charles and Daneen Stiefel Center for Head and Neck Cancer Oropharyngeal Cancer Research Program; and the Program in Image-guided Cancer Therapy.

## Conflicts of Interest

CDF has received related travel, speaker honoraria and/or registration fee waiver from: Elekta AB and unrelated travel, speaker honoraria and/or registration fee waiver from: The American Association for Physicists in Medicine; the University of Alabama-Birmingham; The American Society for Clinical Oncology; The Royal Australian and New Zealand College of Radiologists; The American Society for Radiation Oncology; The Radiological Society of North America; and The European Society for Radiation Oncology. CDF has received related direct industry grant/in-kind support, honoraria, and travel funding from Elekta AB and has served in an unrelated consulting capacity for Varian/Siemens Healthineers. Philips Medical Systems, and Oncospace, Inc.

## 1. Introduction

Adaptive radiation therapy, or radiotherapy (ART), in head and neck cancer (HNC) has shown increasing evidence to benefit patients not only for target coverage and control but also for minimization of normal tissue toxicities[1]. A recent study randomized patients to conventional intensity-modulated radiation therapy (IMRT) with or without weekly adaptation and demonstrated statistically significantly lower rates of xerostomia in the adaptive arm along with corresponding reduction in dose to OARs[2]. Other studies have found that adaptive planning improved median planning target volume (PTV) coverage for doses and yielded significant median dose reductions to key organs-at-risk (OAR) like submandibular glands, parotids, oral cavity, and constrictors[3], led to significant improvements in PTV coverage while reducing maximum dose to surrounding OARs[4], and led to significant improvements in patient quality of life and 2-year local regional control[5].

Looking to the future, phase III randomized clinical trials (RCTs) provide one of the best paths forward towards evidence generation of the benefits of ART[6]. Several phase II RCTs still must be implemented in a phase III setting[7,8], however there are phase III RCTs currently being explored such as the ReSTART (Reducing Salivary Toxicity with Adaptive Radiotherapy) trial[9]. To date, only one phase III RCT has been completed investigating ART in HNC patients which included 132 HNC patients randomized to conventional IMRT with or without weekly adaptations showing no significant differences in primary or secondary endpoints except for parotid gland excretory function[10]. With these phase III trials just beginning now, to move forward towards more effective adaptation schedules, the addition of imaging biomarkers into the current clinical workflow must be addressed now to ensure maximal patient impact. A summary is shown in **Table 1**.

**Table 1:** A summary of completed, on-going, and proposed clinical trials in head and neck cancer with an adaptive radiation therapy component. Abbreviations: CBCT = cone beam computed tomography, CT = computed tomography, DIR = deformable image registration, DLT = dose-limiting toxicity, EFS = event free survival, IMRT = intensity modulated radiation therapy, Linac = linear accelerator, LPFS = locoregional progression-free survival, LRC = loco-regional control, LRFS = locoregional recurrence-free survival, LRR = local regional recurrence, MRI = magnetic resonance imaging, MTD = maximum tolerated dose, OS = overall survival, PCR = pathological complete response, PET = positron emission tomography, PFS = progression-free survival, SBRT = stereotactic body radiation therapy.

| NCT Number (Acronym) | Summary | Study Design | Primary Outcome | Enrollment | Start Date |
|---|---|---|---|---|---|
| **00406289**[11] | Dose escalation in FDG-avid tumor regions | Phase 1, interventional, non-randomized | Tumor recurrence | 24 | Nov 2006 |
| **00608751 (IMRT)** | Adaptive IMRT | Phase 0, interventional | Feasibility | 5 | Jan 2007 |
| **00490282**[12] | Use of CT/MRI to adapt radiation therapy | Interventional | Dosimetric comparison | 25 | June 2007 |
| **01843673** | Use of CT to adapt radiation therapy | Interventional | Dosimetric comparison | 16 | Jan 2009 |
| **01853670 (IGRT)** | Use of CT to monitor delivered dose to patient | Interventional, non-randomized | Clinical patient response | 6 | Aug 2009 |
| **01124409** | Use of CT to guide dosimetric adaptations | Phase 3, interventional, randomized | Early tumor response | 41 | Dec 2009 |
| **02130427** | Use of serial CT/MRI for adaptation | Interventional | Tumor volume | 74 | Sep 2010 |
| **01208883** | Use of serial PET for adaptation | Phase 1, interventional | Volume change | 10 | Sep 2010 |
| **01283178** | Combining IMRT with cisplatin | Phase 1, interventional | Feasibility | 3 | July 2011 |
| **01341535 (C-ART-2)** | Use of serial PET for adaptation | Phase 2, interventional, randomized | Local control | 100 | Aug 2011 |
| **01287390** | Combining IMRT with adaptation per-fx | Phase 2, interventional, randomized | Toxicity | 100 | Oct 2011 |
| **01908504**[13] | Combining PET with CT for adaptive radiation therapy | Interventional | General benefit | 271 | Jan 2012 |
| **01427010** | Use of serial PET for IMRT adaptation | Interventional | Feasibility | 10 | Jan 2012 |
| **02545322 (BART)** | Use of serial CT for adaptation | Interventional | Dosimetric comparison | 18 | Feb 2012 |
| **01504815**[14] **(ARTFORCE)** | Combining PET with CT for adaptive radiation therapy with cisplatin | Phase 3, interventional, randomized | LRFS, toxicity | 268 | Sep 2012 |
| **01874587**[10] **(ARTIX)** | IMRT with weekly replanning | Phase 3, interventional, randomized | Toxicity | 132 | July 2013 |
| **04116047 (CompARE)** | Multi-center dose escalation in adaptive radiation therapy | Phase 3, interventional, randomized | OS, EFS | 785 | July 2015 |
| **02908386 (ROCOCO)** | Use of serial CT for adaptation | Observational | Dosimetric comparison | 0 | Nov 2015 |
| **02653521** | Use of dose tracking, online imaging, and DIR | Interventional, randomized | Toxicity | 80 | Dec 2015 |
| **02952625** | Use of mid-treatment PET/MRI scan for adaptive radiation therapy | Interventional | Image quality, patient tolerance | 8 | Apr 2016 |
| **03096808** | Comparing IMRT with or without adaptation | Phase 2, interventional | LRFS | 64 | Mar 2017 |
| **03215719[15]** | Dose de-escalation from blood-based biomarkers | Phase 2, interventional | PFS | 144 | July 2017 |
| **03286972** | Combining PET/CT with PET/MRI for adaptation | Observational | Examination completion | 3 | Sep 2017 |
| **03376386[16] (ADMIRE)** | Use of PET/CT for dose-escalation mid-treatment | Interventional | Toxicity | 20 | Dec 2017 |
| **04172753[17]** | Feasibility of imaging and treating on an MRI-Linac | Interventional, non-randomized | Feasibility | 472 | May 2018 |
| **03416153** | Use of PET/CT to guide de-escalation mid-treatment | Phase 2, interventional, non-randomized | LRR | 91 | May 2018 |
| **03953352 (GIRAFE)** | Use of serial CT for adaptive radiation therapy | Interventional | Volumetric comparison | 0 | June 2019 |
| **03935672 (PEARL)** | Use of mid-treatment PET/CT for adaptive radiation therapy | Interventional | PFS | 50 | July 2019 |
| **03972072[18] (MARTHA)** | Use of serial MRI on an MRI-Linac for adaptive radiation therapy | Interventional | Toxicity | 49 | Oct 2019 |
| **04242459 (INSIGHT-2)** | Use of MRI for adaptive radiation therapy | Phase 1/2, interventional, non-randomized | Feasibility, MTD | 73 | Oct 2019 |
| **04086901[19] (DART)** | Use of PET for dose escalation | Interventional, randomized | LRC | 3 | Jan 2020 |
| **04188535 (RELAY)** | Use of MRI to assess treatment response | Interventional, non-randomized | Feasibility | 139 | Jan 2020 |
| **04379505 (PEAQ-RT)** | Use of CT for adaptive radiation therapy | Interventional | Feasibility | 10 | Oct 2020 |
| **04612075[20] (EMINENCE)** | Use of PET/MRI for adaptive radiation therapy | Observational | OS | 390 | Jan 2021 |
| **03513042 (EPM-PT-HNSCC)** | Use of PET for adaptive radiation therapy | Observational | LRFS | 12 | Jan 2021 |
| **04477759[21] (DEHART)** | Use of MRI for dose escalation with concurrent atezolizumab | Phase 1, interventional, non-randomized | DLT | 18 | Jan 2021 |
| **04901234 (ART-OPC)** | Use of MRI for mid-treatment adaptive radiation therapy | Phase 2, interventional, randomized | Toxicity | 120 | July 2021 |
| **05081531 (RadiomicArt)** | Use of CT, MRI, and PET for adaptive radiation therapy | Interventional | LRFS | 50 | Oct 2021 |
| **05160714 (MRL-02)** | Feasibility of MRI for dose response in adaptive therapy | Phase 1, interventional | DLT | 24 | Jan 2022 |
| **04883281 (DARTBOARD)** | Potential benefit of daily adaptive radiation therapy | Phase 2, interventional, randomized | Toxicity | 50 | Feb 2022 |
| **05348486 (FARHEAD)** | Detection of radioresistance using PET/CT for dose escalation | Interventional, non-randomized | LPFS | 120 | Apr 2022 |
| **05393297 (InGReS)** | Use of MRI and PET/CT for dose escalation combined with chemotherapy | Interventional | Toxicity | 15 | June 2022 |
| **05831917** | Use of MRI-Linac for dose reduction. | Interventional | Toxicity | 41 | Jan 2023 |
| **05996432** | Use of MRI for identifying radioresistence | Phase 0, interventional, non-randomized | Imaging biomarkers | 48 | May 2023 |
| **04809792** | Feasibility of MRI-Linac for serial SBRT | Interventional | Feasibility | 30 | June 2023 |
| **05666193** | Use of HyperSight CT for adaptive radiation therapy | Interventional | Dosimetric comparison | 30 | July 2023 |
| **05919290 (HN-Quest)** | Use of MRI biomarkers for adaptive radiation therapy | Interventional, non-randomized | Toxicity | 173 | July 2023 |
| **06691776 (FASCINATE)** | Feasibility of CBCT for adaptive radiation therapy | Interventional, non-randomized | Feasibility | 100 | Aug 2023 |
| **06116019 (ART-02)** | Use of CBCT for adaptive radiation therapy | Observational | Feasibility | 649 | Oct 2023 |
| **06214611 (ART in HNT)** | Identify early markers of benefit from adaptive radiation therapy | Interventional, non-randomized | Dosimetric comparison | 50 | Nov 2023 |
| **06137274** | Use of MRI for adaptive radiation therapy | Interventional | Volumetric comparison | 25 | Dec 2023 |
| **06005324** | Use of blood-based biomarkers for adaptive radiation therapy | Phase 1, interventional, non-randomized | Blood-based biomarkers | 36 | Dec 2023 |
| **06234748 (ARTHOUSE)** | Use of MRI and PET for dose escalation | Phase 2, interventional | Toxicity | 19 | Dec 2023 |
| **06345287** | Develop adaptive radiation therapy plan with induction immuno-/chemo-therapy | Phase 2, interventional, non-randomized | PFS | 133 | Jan 2024 |
| **06041555 (ISRAR)** | Use of MRI for adaptive radiation therapy | Observational | Imaging biomarkers | 600 | Jan 2024 |
| **06216171 (ProHEART)** | Use of imaging for anatomically optimized adaptive radiation therapy | Interventional, randomized | Toxicity | 30 | Jan 2024 |
| **06323460**[22] | Use of blood-based biomarkers for dose de-escalation | Phase 2, interventional, non-randomized | OS, PFS | 45 | Mar 2024 |
| **06361043**[23] **(ARTEC)** | Evaluate toxicity in adaptive radiation therapy patients | Observational | Toxicity | 30 | Apr 2024 |
| **06516133 (OART)** | Use of adaptive radiation therapy in minimizing toxicities | Phase 3, interventional, randomized | LRFS | 494 | May 2024 |
| **06572423**[24] **(PULS-Pal)** | Use of dose escalation with HyperArc | Interventional | PFS | 43 | Oct 2024 |
| **05849142**[25] **(OPC-V)** | Use of MRI-Linac for adaptive radiation therapy | Interventional | Dosimetric comparison | 0 | Dec 2024 |
| **06446713 (PIRATES)** | Feasibility of imaging for proton dose escalation | Phase 1, interventional | Feasibility | 17 | May 2025 |
| **06990178** | Feasibility of low-dose adaptive radiation therapy | Interventional | PCR | 43 | Mar 2028 |

However, the evolving landscape of biomarker development in medical research demands standardized approaches and clear operational definitions. Traditionally, biomarkers were primarily understood as biological indicators measured and evaluated to assess normal biological processes, pathogenic processes, or pharmacologic responses to therapeutic interventions. However, the FDA-NIH Biomarker Working Group[26] has since expanded this definition to explicitly include molecular, histologic, radiographic, and physiologic characteristics as types of biomarkers and defined specific biomarker categories based on their context of use which have translational potential to radiation oncology (e.g., monitoring or predictive biomarkers in the context of radiation-induced toxicities).

Despite this, there is still a translational gap that needs to be addressed so that research-based tools and knowledge can be integrated into clinical pathways for improved routine patient care in cancer[27]. To facilitate and support this translation for potential clinical interventions or decision-making, clinical trial-based frameworks have been suggested (e.g., van Houdt et al.[28] and Boss et al.[29]). On the other hand, validation of image biomarkers should also be addressed from a more technical standpoint such as with DECIDE-AI[30,31], R-IDEAL[32], and radiomics[33]. These have already been adopted in multiple studies[34], however, most efforts in standardization of use and reporting of image biomarkers (e.g., the Quantitative Imaging Biomarkers Alliance, or QIBA) has been driven by radiology-related professional bodies which may lack full insights or understanding of the radiation oncology specific applications and needs. While diagnostic applications rely largely on qualitative use of imaging, radiation oncology often requires a more quantitative approach to facilitate clinical decision-making based on a threshold or cut-off for the continuous quantitative image biomarker data[27,35]. There have been considerable technical advances in imaging space that allow for these imaging biomarkers to exist and enable longitudinal assessments, thus supporting clinical decision making in the ART setting. While current technology in ART is generating positive studies for its benefit compared to conventional methods, technical improvements just around the corner and in line for future implementation have the potential to enhance these benefits. This review article aims to provide an overview of these technical advances and the status of ART with insights on clinical translation and applications in the HNC setting.

## 2. Current Practice

### *2.1. General Image Guided Radiation Therapy (IGRT)*

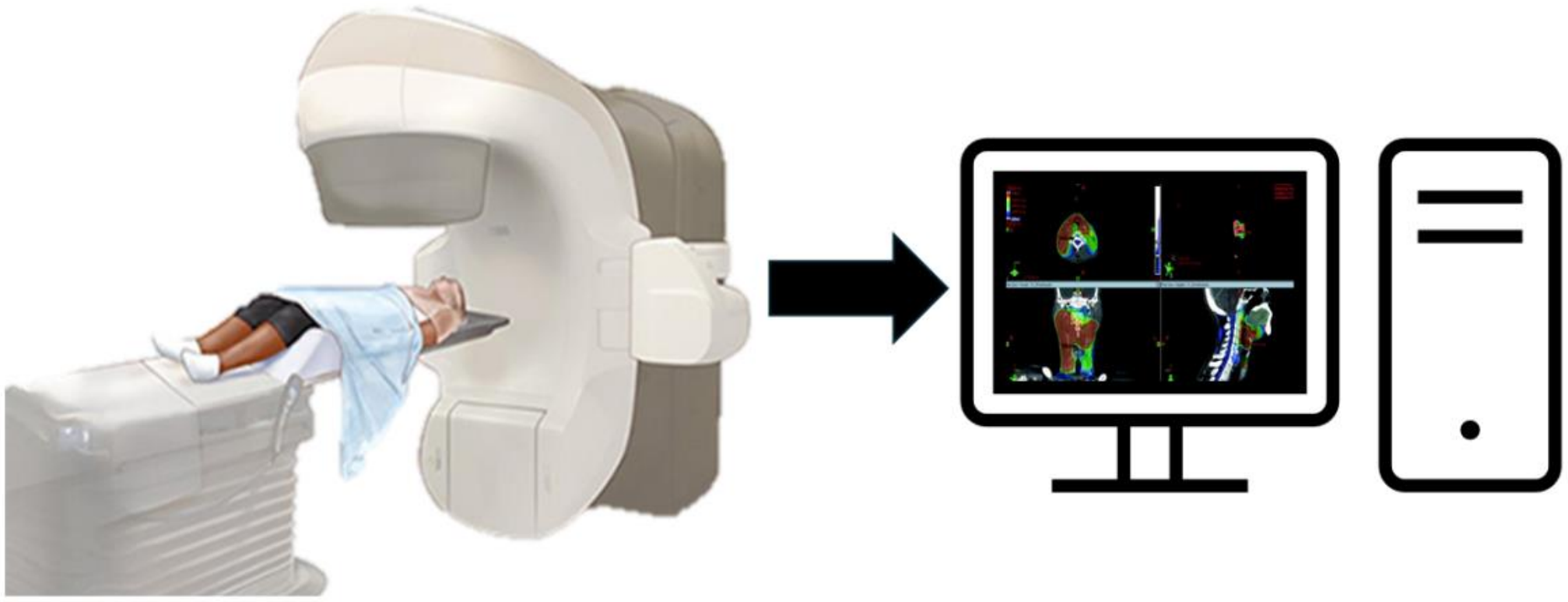

General image guided radiation therapy (IGRT) is a longstanding technique that has undergone continued advancement used to improve the precision and accuracy of treatment. It refers to the use of various imaging modalities, such as X-ray, computed tomography (CT), or magnetic resonance imaging (MRI) to visualize the tumor and surrounding tissues both during patient setup and treatment delivery. This ensures that the dose is administered to the patient as planned[36]. By imaging the patient's anatomy in the treatment position for reference, we can reduce geometric positioning errors and ensure that radiation is delivered directly at the target while avoiding normal tissue[37]. Adaptive IGRT has emerged to allow for real-time plan adjustment depending on threshold changes such as changes in tumor volume[38]. In summary, IGRT technology is crucial to the delivery of safe and effective ART, and radiation therapy in general[39].

### *2.2. Image Guided Adaptive Radiation Therapy*

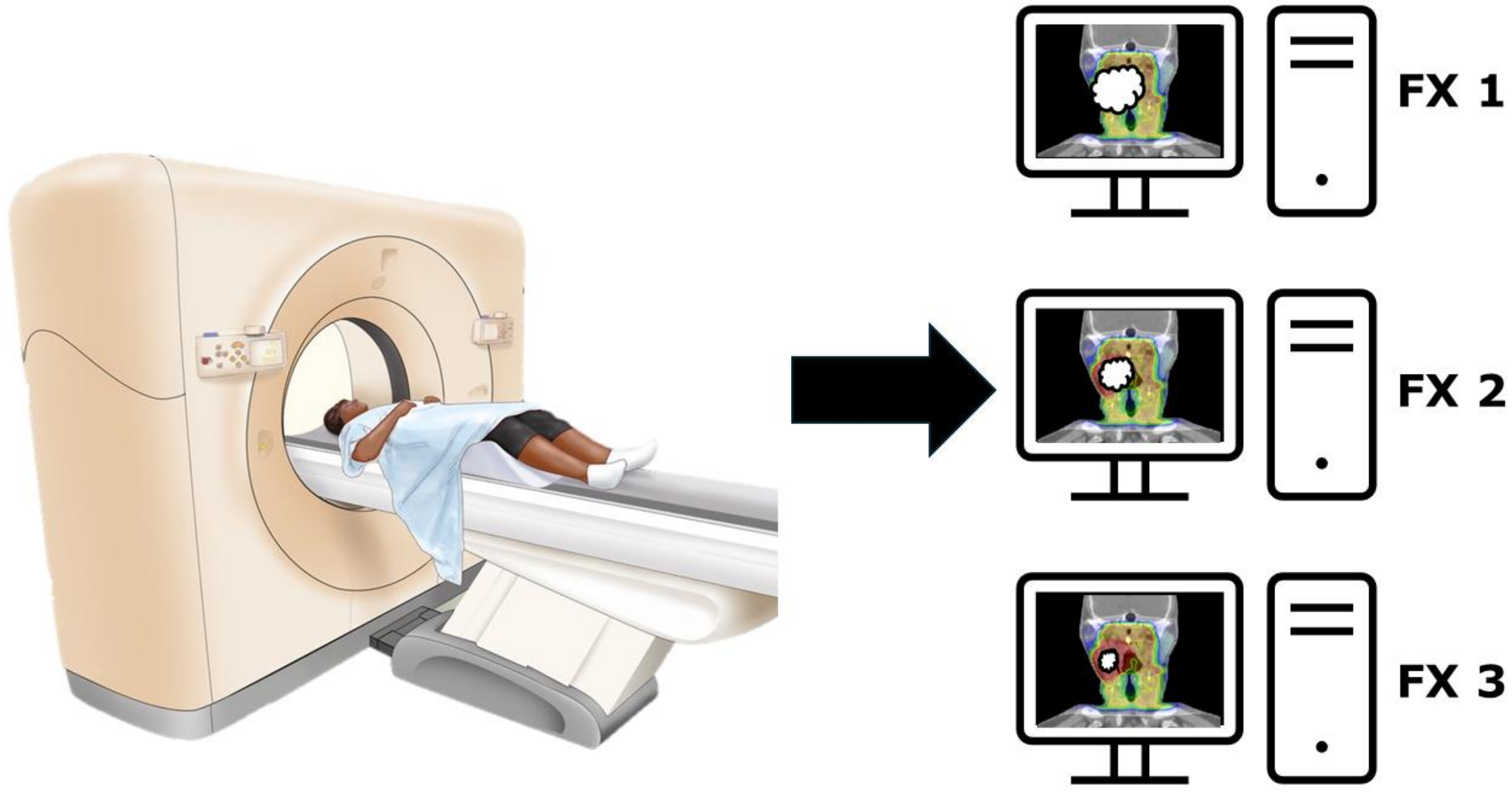

### 2.2.1. Halcyon

The Halcyon system has emerged in radiotherapy in the treatment of head and neck cancers (HNC) specifically, offering a streamlined and patient-friendly approach[40]. This platform, which integrates advancements in linear accelerator technology, is designed to deliver high-quality treatment with efficiency and precision. One aspect of Halcyon's current application in head and neck radiotherapy is its capability for IGRT, which allows for better tumor targeting while minimizing radiation exposure to surrounding healthy tissues. The integration of IGRT with Halcyon's simplified workflow enhances patient throughput and reduces the time patients spend on the treatment table, which is beneficial for those with HNC[41].

Another advantage of the Halcyon system is its use of volumetric modulated arc therapy (VMAT), which enables highly conformal dose distributions with rapid treatment delivery[42]. This is especially pertinent in HNC cases, where the proximity of critical structures such as the spinal cord and salivary glands requires meticulous planning and delivery[43]. The Halcyon system's dual-layer multi-leaf collimator (MLC) is specifically designed to optimize VMAT treatments, ensuring sharp dose gradients that spare normal tissues without compromising tumor control[44]. Clinical studies have demonstrated that Halcyon-based VMAT can achieve comparable or superior outcomes in head and neck radiotherapy compared to traditional linear accelerators, with a notable reduction in treatment times[45].

Furthermore, Halcyon's ability to operate within a smaller footprint and with reduced operational noise has made it a preferred choice in modern radiotherapy departments. Its design not only enhances patient comfort but also aligns with the evolving demands of healthcare facilities seeking to optimize space and resources. The streamlined treatment planning and delivery process facilitated by Halcyon allows clinicians to focus more on patient care and less on managing the complexities often associated with traditional radiotherapy systems. As a result, the adoption of Halcyon is increasingly seen as a best practice for head and neck radiotherapy, with ongoing research and clinical practice continually refining its application[46,47].

### 2.2.2. CT-on-Rails

CT-on-Rails is an IGRT technology that integrates a dedicated CT scanner into the radiotherapy workflow. This system provides significant advantages for ART in the treatment of HNC, where precise targeting is crucial due to the complex anatomy and the proximity of critical structures[48].

The CT-on-rail system is mounted on a moveable framework that can be positioned around the patient to acquire high-quality, high-resolution images before or during treatment[49]. This allows for daily image guidance, essential in head and cancer treatments, given the potential for significant anatomical changes over radiation therapy. These changes may include tumor shrinkage, patient weight loss, or movement of critical structures, all of which can affect the accuracy of radiation delivery[50].

ART with CT-on-rails involves using daily CT images to monitor anatomical changes and adapt the treatment plan accordingly. This approach can improve treatment accuracy and potentially enhance outcomes by radiation dose conforms closely to the tumor while sparing healthy tissues. In other words, if the tumor shrinks significantly, the treatment plan can be adjusted to avoid over-irradiating normal tissues[48].

CT-on-rails systems provide the flexibility of conducting imaging and treatment positioning separately, which can be beneficial in certain clinical workflows. Although cone-beam CT (CBCT) is commonly used for patient setup due to its integration into radiotherapy systems and the ability to perform imaging directly in the treatment position, CT-on-rails offers several unique advantages that can be particularly beneficial in specific clinical situations. Although cone-beam CT (CBCT) is commonly used for patient setup due to its integration into radiotherapy systems and the ability to perform imaging directly in the treatment position[51], CT-on-rails offers unique advantages. This reduces the need for repositioning the patient multiple times, thereby minimizing the overall impact on treatment despite the initial separate imaging and positioning steps[48].

Clinical studies have demonstrated the effectiveness of CT-on-rails in HNC radiotherapy[48,52]. Further, it has been shown that the technology can significantly improve tumor control and reduce toxicities compared to on-adaptive approaches[48]. By facilitating frequent and precise imaging, CT-on-rails enables personalized treatment adjustments that account for individual patient anatomy and tumor response. This individualized approach is increasingly seen as a gold standard in HNC radiotherapy, aligning with the broader trend toward personalized medicine in oncology[53]. As technology continues to evolve, the role of CT-on-rails in ART is likely to expand, further improving outcomes for patients with HNC.

### *2.2.3. Ethos*

The Ethos™ radiation therapy system (Varian Medical Systems, Palo Alto, CA, USA) is a form of online ART utilizing CBCT-guided ART alongside artificial intelligence (AI) and machine learning[54]. At each visit, a new planning image is taken using CBCT and is linked to the original image, generating a synthetic CT. This CT is then automatically segmented using AI or deformable image registration (DIR). Using the Ethos™ intelligent optimization engine (IOE), an adaptive treatment plan is created. The initial treatment plan is also generated using the IOE[55]. The Ethos™ system can be advantageous over conventional CBCT, as it uses image reconstruction to reduce previous issues such as high scatter and poor image quality[54]. Ethos™ has been evaluated in the setting of numerous cancers including prostate, breast, rectal, as well as metastases in an immunostimulatory low-dose radiation setting[56–60].

The use of Ethos™ in HNC has also been evaluated. Multiple retrospective studies highlight the acceptability of Ethos™ generated plans[55,61]. El-Gmache and McLellan selected 10 previous patients and generated IMRT and VMAT plans using Ethos™ for each patient. Plans were reviewed by consultants, with most being considered acceptable, with IMRT being rated higher than VMAT plans[55]. Other studies report that the use of Ethos™ is feasible in HNC[3], demonstrating satisfactory contours subjectively by physician review and objectively using dice similarity coefficient[62], as well as statistical improvement over scheduled plans in terms of coverage and benefits to certain organs-at-risk (OAR) such as the larynx, parotid, brainstem, and spinal cord[63]. Similarly, ART with Ethos™ shows potential in decreasing dose to OARs in tongue and tongue base tumors[64]. In summary, Ethos™ has been shown to be an effective option in ART for HNC and continued use in the future is expected.

#### *2.2.4. MR-Linac*

Magnetic resonance-guided linear accelerators (MR-Linac) combine an MRI scanner with a linear accelerator. The development of the MR-Linac has been considered a significant advancement in ART[65]. The advantages of using MRI in image guidance over other methods such as CT include higher soft tissue contrast and the lack of ionizing radiation. CT-based methods also have issues involving poorer image quality and high scattering[66]. An MR-Linac, developed by Elekta and Phillips, known as the Elekta-Unity system is currently commercially available, and was first used clinically in 2018 for oligometastatic lymph nodes[67]. The goal of the MR-Linac system is to observe and account for anatomical changes in real-time over the course of a patient's treatment, constantly adapting the treatment plan to fit this. As highlighted by Ng et al., there are three major advances in ART performed by the MR-Linac system[65]. These include imaging for therapy guidance, adaptive treatment planning for inter-fractional management, and real-time imaging and gating for intra-fractional management.

This system's clinical application has been studied in several sites and cancers, including chest/lung, abdominal, and genitourinary tumors[65,66,68–70]. The MR-Linac has also been evaluated in the setting of HNC[66,71]. MRI imaging techniques are useful in HNC due to its advantage in soft-tissue imaging. Additionally, HNC regresses quickly, due to their high sensitivity to treatment. Due to this, adaptive therapy can be utilized, and MR-Linac systems are an efficient way to do this[72]. MRI-guided radiation therapy has been shown to minimize adverse effects and reduce dosages to OARs[66]. However, with the use of MR-Linac, dose accumulation is a consideration that must be made. Since MR-Linac systems adapt by fraction, rather than total dose, there can be issues in understanding the total delivered dose. McDonald et al. offer steps in addressing the issue of dose accumulation[73]. Overall, the MR-Linac is a valuable tool in effective ART adaptation and delivery.

### *2.3. AI-Based Applications*

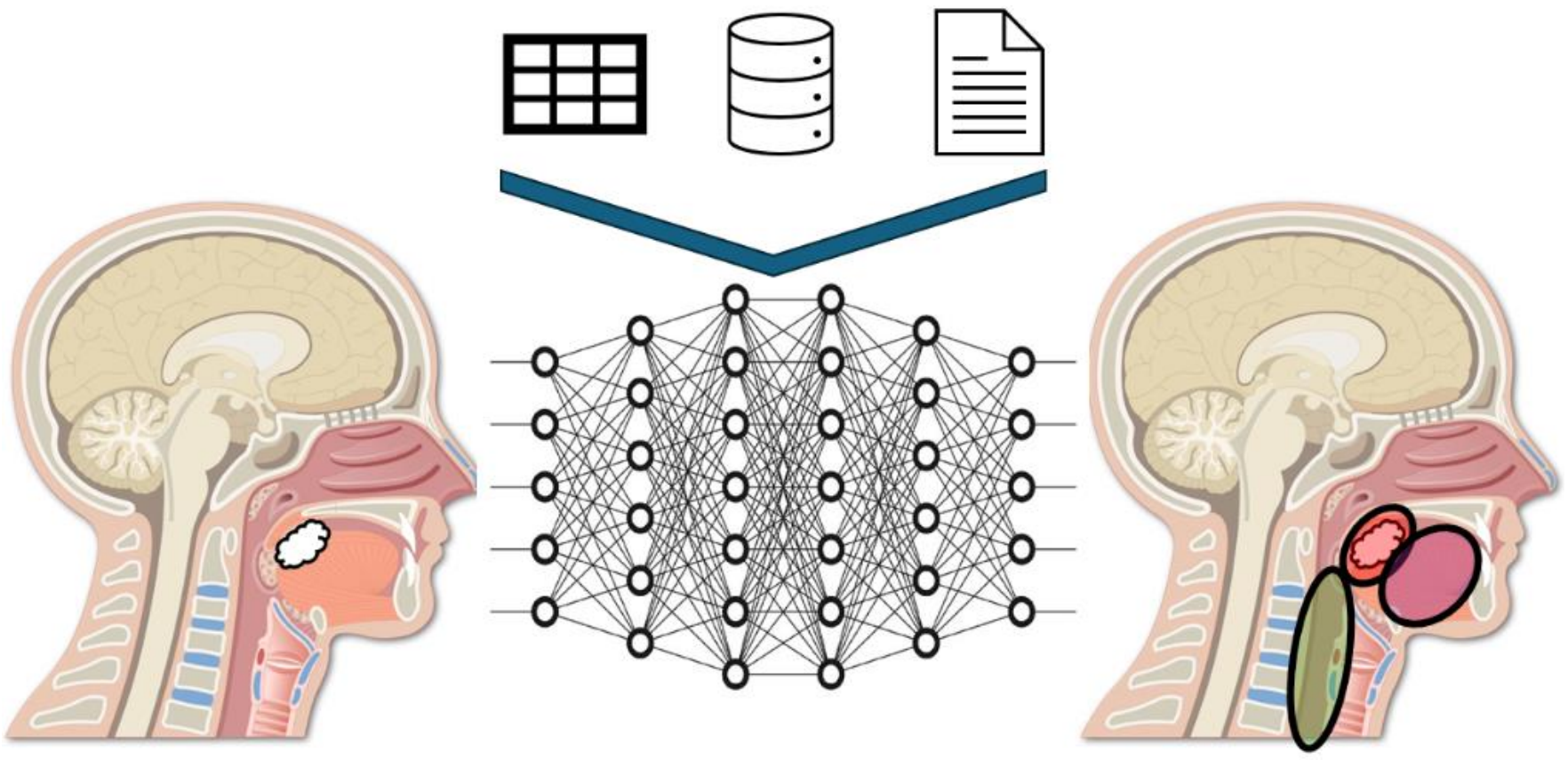

AI-based segmentations have become a significant innovation in medical imaging, particularly in tasks like delineating organs and tumors in CT, MRI, and positron emission tomography (PET) scans[74–76]. Unlike traditional manual methods, which are time-consuming and prone to variability, AI-driven approaches offer automation, accuracy, and consistency[77].

Deep learning models, especially Convolutional Neural Networks (CNNs) and U-Nets, are at the forefront of this advancement. These models are trained on large datasets to identify and segment specific structures in medical images, improving efficiency in clinical workflows like radiotherapy planning. Studies have shown that AI can reduce the time required for tasks like contouring OARs while maintaining accuracy comparable to expert clinicians[78,79].

However, there are challenges. AI models often require large, annotated datasets for training, which is time and resource intensive. Additionally, most models currently do not provide an uncertainty quantification score to their predictions limiting their adoption into clinical practice due to mistrust[80]. Further, differences in imaging protocols and patient populations can affect how well these models perform in different clinical settings. Techniques like transfer learning and domain adaptation are being explored to improve generalization[74,81].

Despite these hurdles, the potential of AI-based segmentation is vast; as this technology evolves, it will play an increasingly vital role in improving the precision and efficiency of medical imaging, specifically in the ART setting.

## 3. Around the Corner

### *3.1. PET-Guided ART: RefleXion*

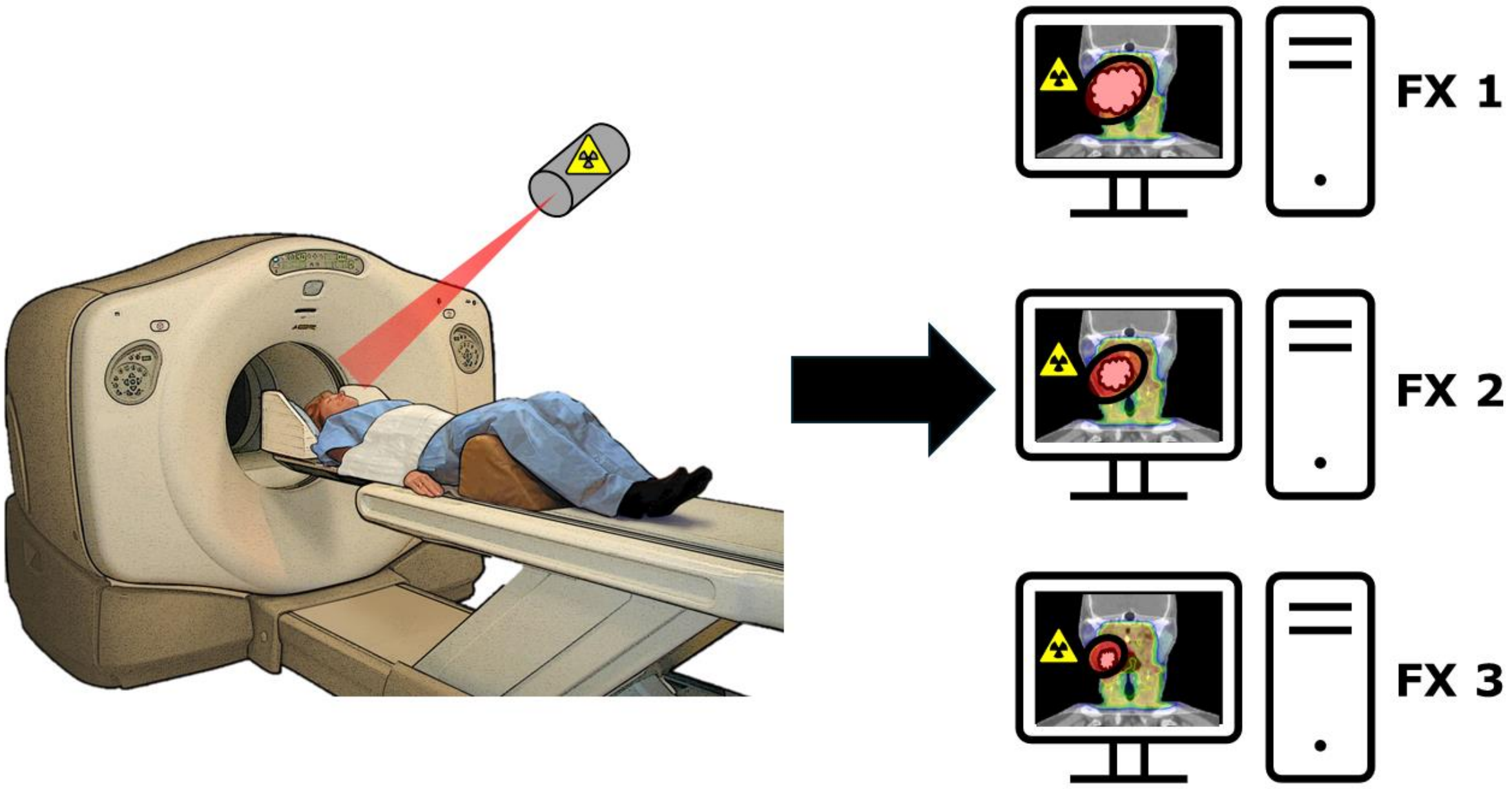

Recently, a novel device combining a PET and CT system, named RefleXion[82], has been introduced by the company RefleXion Medical Inc. (Hayward, CA, USA). This machine requires its own treatment planning system which has been successfully built and commissioned on head and lung phantoms[83,84]. In addition, the real-time delivery system has been tested across static and moving targets in phantoms and shown to have gamma passing rates upwards of 95%, making it clinically feasible[85]. Technical validation of the on-board PET imaging arrays has been conducted with comparable spatial resolution and image contrast as typical diagnostic PET scanners[86]. The treatment delivery system has also shown similar performance in phantoms[87] and in HNC patients[88]. Further, new treatment workflows have been introduced for this novel device, however more rigorous technical reports should be created as this system is deployed to more clinics[89].

One study described the deployment and application of this machine for both stereotactic body radiation therapy (SBRT) and intensity modulated radiation therapy (IMRT) treatments primarily in the head and neck at 63% of all treated sites[90,91]. Further, in the head and neck, novel tracers such as $^{89}$Zr-panitumumab have been combined with traditional tracers like $^{18}$F-FDG for detection and staging of head and neck squamous cell carcinoma[92] (HNSCC). Additionally, $^{68}$Ga-DOTATATE has been shown to be useful in the diagnosis and management of HNC, making it another potential application of the RefleXion for ART[93].

### *3.2. Real-Time Motion Monitoring on the MR-Linac*

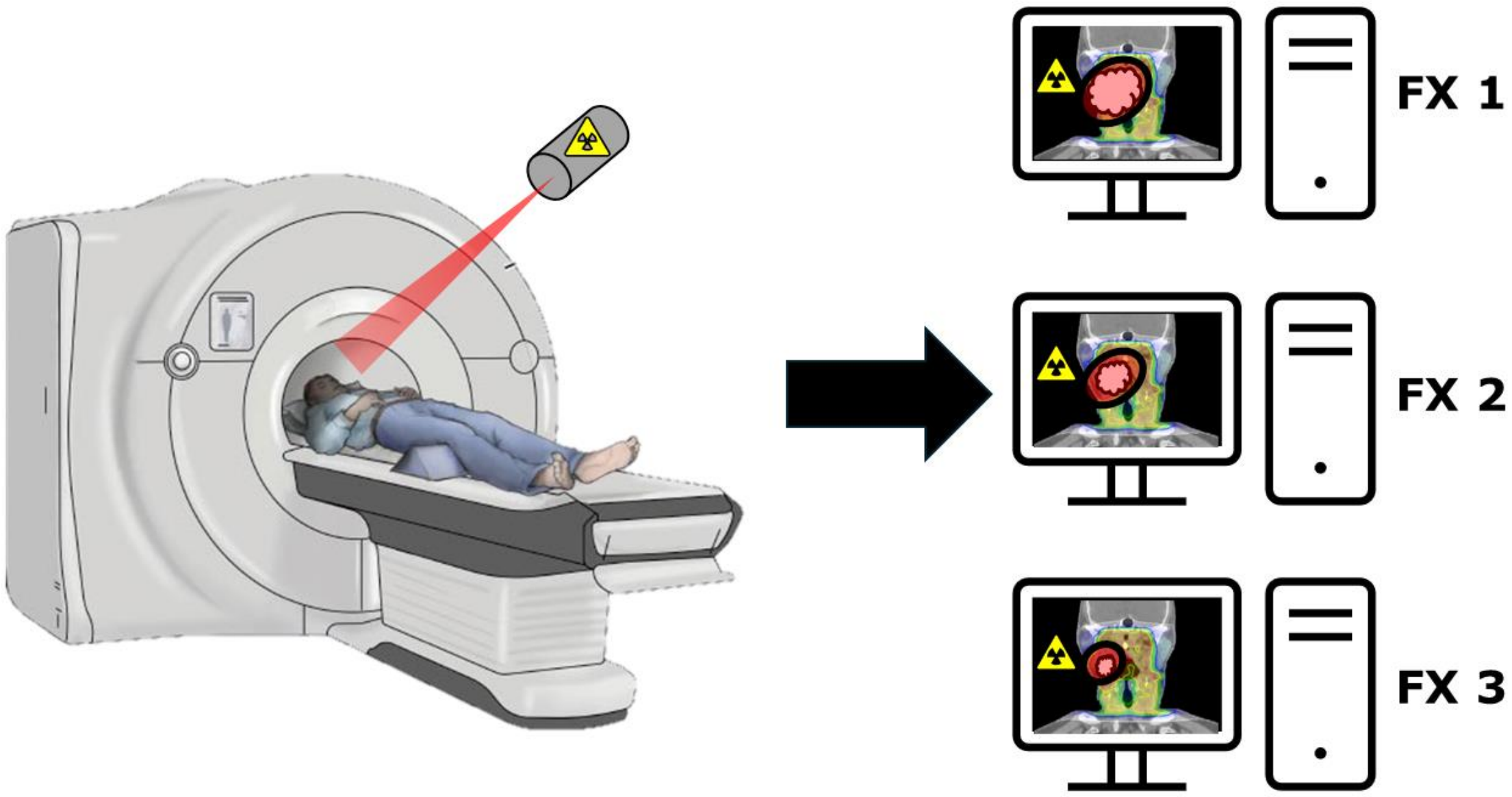

One of the promises of the MR-Linac is the real-time tracking of region-of-interest motion and adapt the treatment based on those changes[66,94]. This feature has been applied to the ViewRay (Oakwood Village, OH, USA) MRIdian 0.35T MR-Linac[65] and recently been introduced by Elekta AB for the Unity 1.5T MR-Linac[95] and known as the comprehensive motion management (CMM) system. Several treatment strategies are available with this software: (1) free-breathing exhale where the treatment beam is paused when the target moves outside the exhale position, (2) free-breathing average position which only pauses the beam when the target moves outside a pre-defined expected path, (3) breath hold when the beam is paused outside of the breath hold position, and (4) expectation based where the beam is paused when the motion exceeds a pre-defined margin. Each technique has its own set of advantages and limitations which should be strictly considered when deciding optimal ART strategies for the head and neck.

This technique is particularly attractive for the upper airway of HNC patients who experience the most motion due to swallowing and related tongue adjustments throughout the treatment session. Motion in these areas of greater than 5 mm and 10 mm was seen in 13% and 4% of total imaging time intrafraction and 24% and 3% of total imaging time interfraction, respectively[96]. Cine-based MRI sequences have been developed by Paulson et al. in 2011 for assessment of anatomical changes in HNC patients[97]. Additional validation was performed by Bradley et al. in 2011 using a similar approach[98]. Further, a model-driven method for tracking these anatomical changes has also been developed[99].

### *3.3. MRI-Based Diffusion: Diffusion Weighted Imaging (DWI) and Apparent Diffusion Coefficient (ADC)*

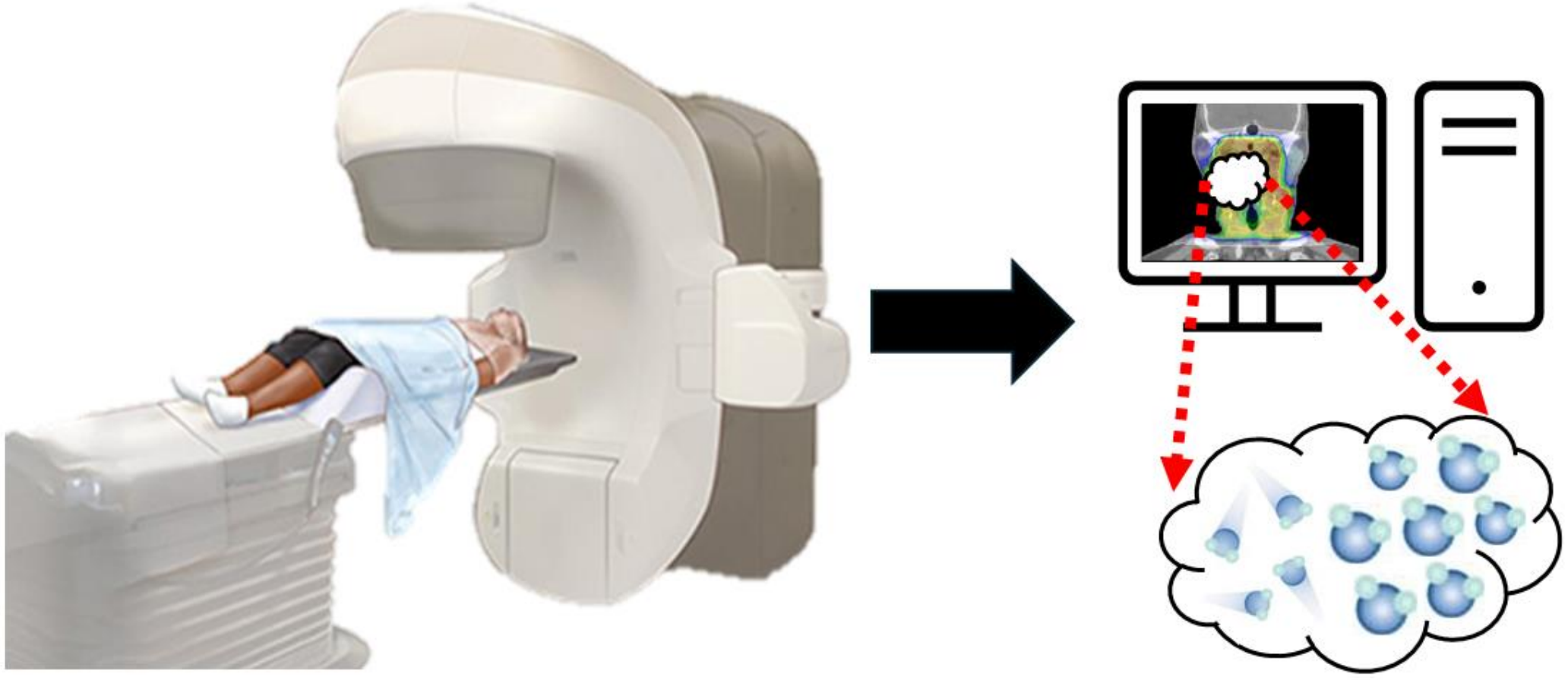

Measuring diffusion changes in normal and malignant tissues is critical to understanding ongoing underlying damage mechanisms and tumor progression making it a strong contender for use in ART. The brief technical overview of extraction of diffusion information from MRI is that protons inside the body can be stimulated with a radiofrequency pulse and allowed to move around for a short period of time before an equal and opposite refocusing pulse is applied to bring the signal back to baseline[100,101]. In the case of non-diffusing protons, the signal is completely cancelled, however in the case of diffusing protons, the signal will not be perfectly cancelled due to the varying frequency encoding across the imaged area. The strength of the gradients used to encode the frequency information is often expressed in terms of the b-value which determines the diffusion sensitivity of the resulting image. These gradients can also be oriented in any direction for a desired diffusion measurement or combined resulting in diffusion tensor imaging (DTI).

ADC as a biomarker for adaptive treatment in the head and neck has been prospectively validated across 81 patients finding significant increases at mid-RT compared to baseline in those with complete response while those without complete response measured no significant differences[102]. As a result, on the MR-Linac, the repeatability and reproducibility of ADC has been tested in phantoms and in-vivo and found within-subject coefficients of variation under 10% and 12%, respectively in tumors across varying DWI acquisition techniques[103,104]. Additional validation of ADC as a biomarker has recently been investigated and showed a significant negative correlation of mean change in ADC with the change in the volume of the gross primary disease[105–108]. In response, the MR-Linac Consortium has meticulously detailed the recommendations for measuring ADC on the MR-Linac[109].

### *3.4. MR-Only Workflow on the MR-Linac*

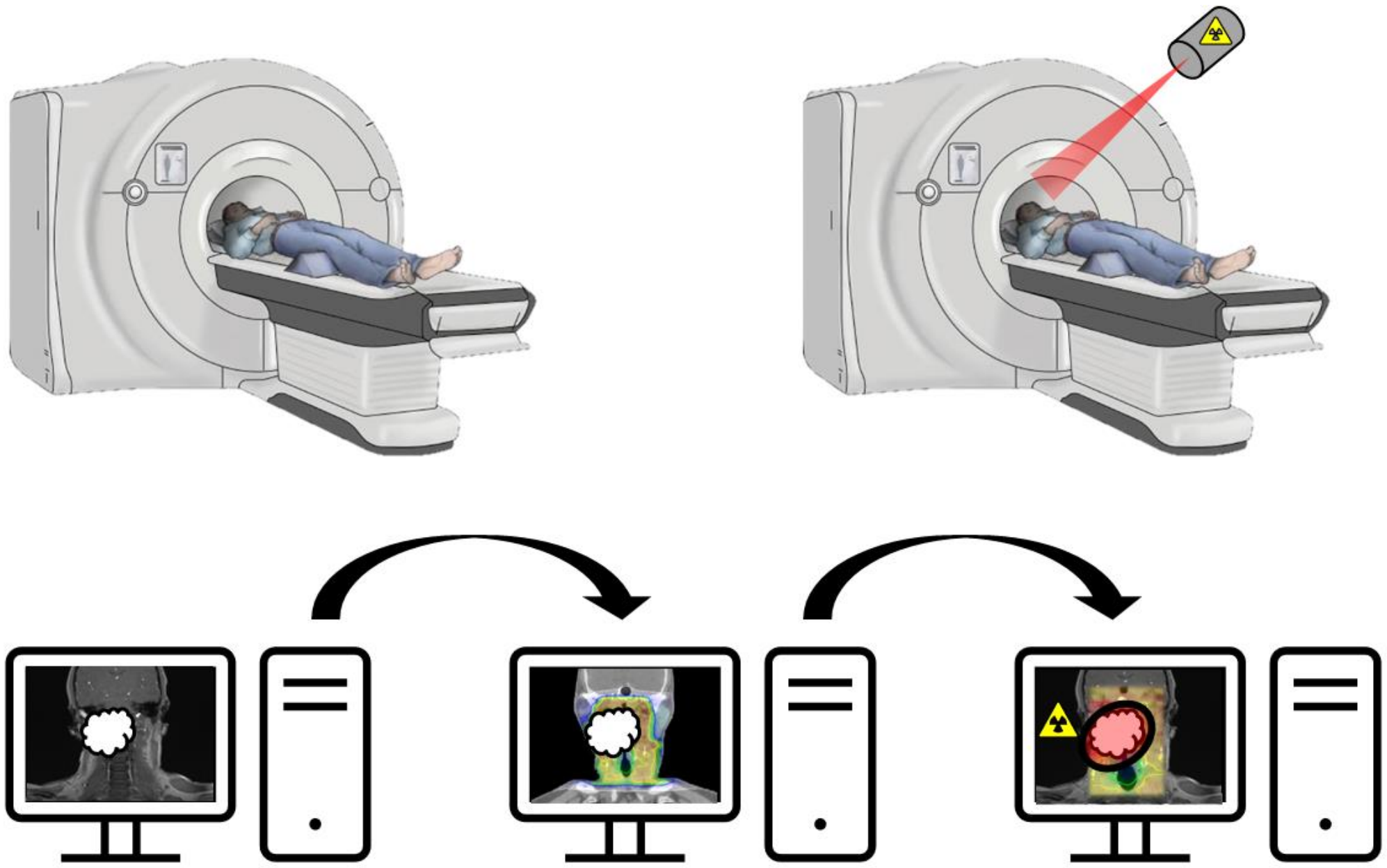

One major challenge of the MR-Linac is that the treatment is conducted based on MRI changes, however electron density values are required from the CT to accurately determine the expected dose distributions[110]. This creates inherent registration issues between the modalities and potential for error propagation throughout a long MR-Linac fractionation scheme, which is typical for HNC[111]. Therefore, extensive research investigating the potential to extract the required CT information from MRI scans synthetically has been conducted.

One approach is to segment the bony anatomy in MRI scans and use this information to generate dictionary-based mapping between MRI signal intensity and CT number ranges within the regions of interest[112]. This information can then be used to create synthetic digitally reconstructed radiographs (DRR) or pseudo-CT. Another popular approach is to use reference atlases (i.e., multi-atlas algorithm) generated from registered single patient MRI and CT scans to create a mapping algorithm between the two modalities[113]. With the rise of AI-based solutions, respective translations to the synthetic-CT problem have been plentiful using approaches such as convolutional DenseNets[114], patch-based 3D CNNs[115], patch-based generative adversarial neural network (GAN) models[116], 3D deep CNNs[117], compensation cycle consistent GANs[118], and structure completion GANs[119]. Multiparametric approaches have also been proposed which utilize information from T1-weighted, T1-weighted post-contrast, T1 Dixon post-contrast, and T2-weighted images, however compounding registration errors are still a concern[120].

Besides algorithms for synthetic-CT generation, workflow optimizations have been suggested such as the implementation of immobilization masks, typically reserved for the radiation therapy applications, to also be applied during scans on diagnostic imaging systems for enhanced comparisons[121]. Pilot studies evaluating the potential of current synthetic-CT algorithms have been tested in the palliative setting across several disease sites including head and neck[122] and more rigorously in the pelvic region due to its simple and homogeneous anatomy[123]. However, more work will have to be done before routine clinical implementation due to its respective complexity.

### *3.5. Dynamic Contrast Enhanced (DCE) MRI*

Dynamic contrast enhanced (DCE) MRI, also known as permeability MRI, is considered one of the essential MRI techniques and has been widely used clinically for its non-invasive techniques for improved detection and monitoring of diseases. DCE MRI analyzes how a tissue's signal changes over time using an intravenously injected paramagnetic contrast agent. After a pre-contrast baseline image is acquired, a temporal series of T1-weighted images are run to capture the flow of contrast throughout the bloodstream[124,125]. Due to its decreased spatial resolution, DCE MRI is often paired with anatomical MRI sequences (T1-weighted, T2-weighted). These images can be registered for enhanced functional and anatomical information at localized space- and timepoints. DCE MRI can be used to help differentiate benign and malignant tissue and to identify vascular abnormalities, such as arteriovenous malformations[126,127]. Additionally, DCE MRI has recently been demonstrated as a determinant in early detection of vascular changes in normal tissue following radiation therapy in the head and neck[128].

The technical process of DCE-MRI utilizes specialized imaging sequences and contrast agents to capture the dynamic changes in tissue enhancement over time. Typically, two primary approaches are utilized: MR angiography-type sequences, which focus on evaluating blood flow dynamic, and fast gradient-echo sequences, which offer greater anatomical resolution. To achieve the necessary balance between temporal and spatial resolution, parallel imaging techniques and phased array coils are essential. This makes it possible to conduct in-depth analyses of vascular characteristics and tissue perfusion in various clinical settings, including oncology and cardiovascular imaging as shown in **Figure 1**.

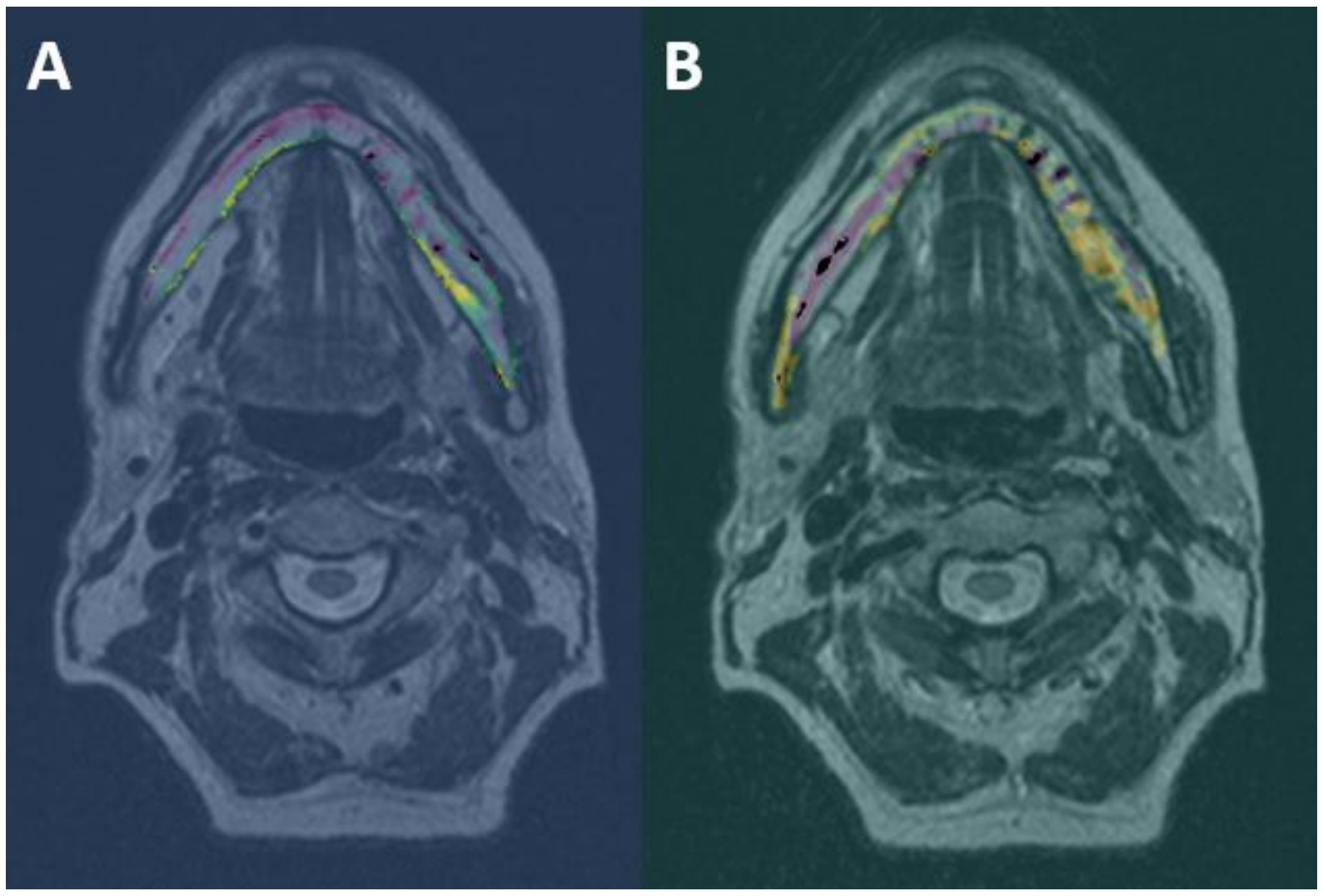

**Figure 1:** $\Delta K_{trans}$ maps (A) and $\Delta V_p$ maps (B) of the mandible derived from DCE-MRI overlayed on registered T2-weighted images, indicating local vascular changes.

### *3.6. Chemical Exchange Saturation Transfer (CEST) MRI*

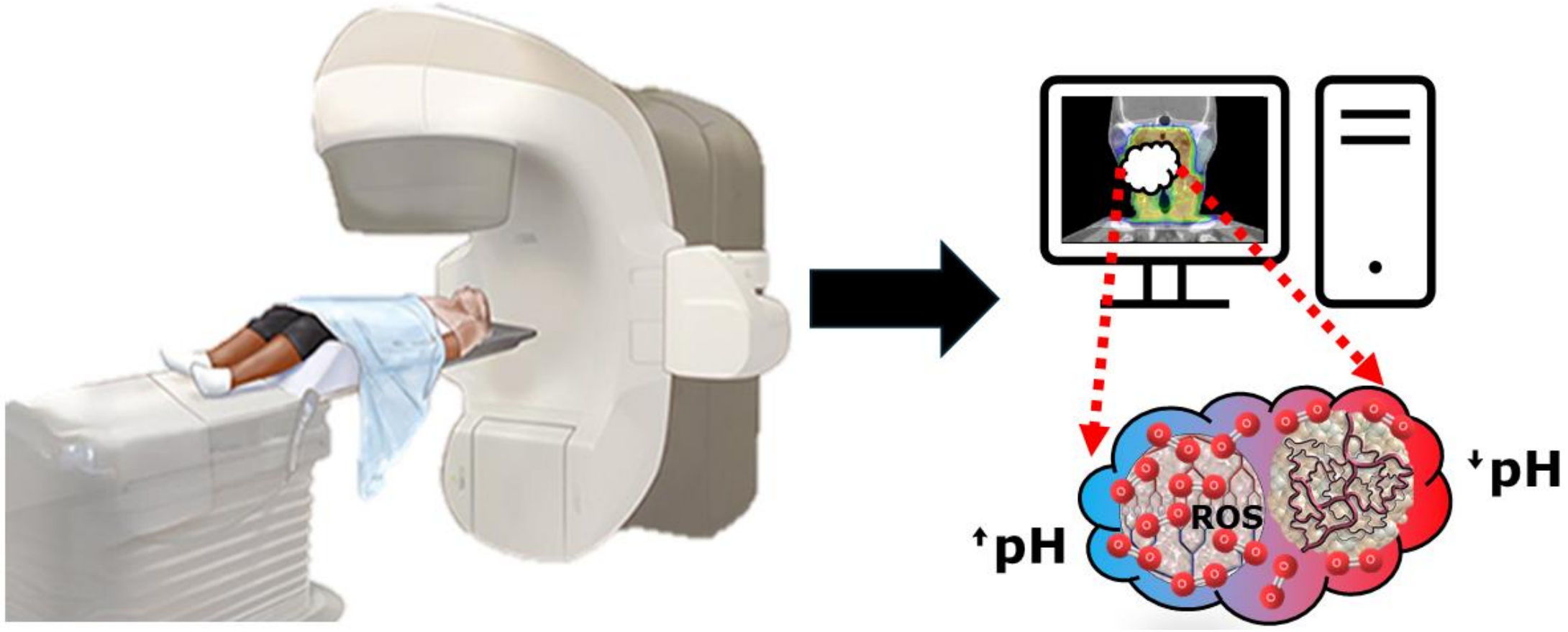

Chemical exchange saturation transfer (CEST) is a versatile technique with a wide range of assessable biomarkers. CEST techniques vary significantly, from evaluating transferred saturation via the Nuclear Overhauser Effect (NOE) to utilizing exogenous agents to saturating non-1H nuclei such as 19F. Several techniques are of especial interest to radiation therapy. A combination perfluorooctyl bromide and glycerol nano-molecular imaging probe has been developed for dual pH and O2 sensing via 19F/1H-CEST[129]. This has not yet been assessed clinically but has clear potential in assessment prior to radiation therapy. Glucose CEST (glucoCEST) has also been used previously in HNC patients on a 3T scanner demonstrating significantly higher signal enhancement in tumors versus normal muscle[130].

Amide proton transfer (APT) is the most prolific subvariety of CEST due to comparative ease of application. APT uses CEST-active amide protons within endogenous proteins negating the need of an injectable agent and streamlining acquisition. This has been applied to predict short-term therapeutic outcome in nasopharyngeal carcinoma[131] and to differentiate high versus low grade CNS tumors on a 1.5T MR-Linac[132]. The primary application of APT-CEST regarding radiation oncology is in differentiating radiation necrosis from tumor recurrence followed by treatment response in CNS tumors[133–141]. Combination studies utilizing several of these described techniques have been used, demonstrating improved power using a multiparametric approach[142–147] which will be discussed further in the next section.

## <u>4. Future Directions</u>

### *4.1. T1ρ MRI*

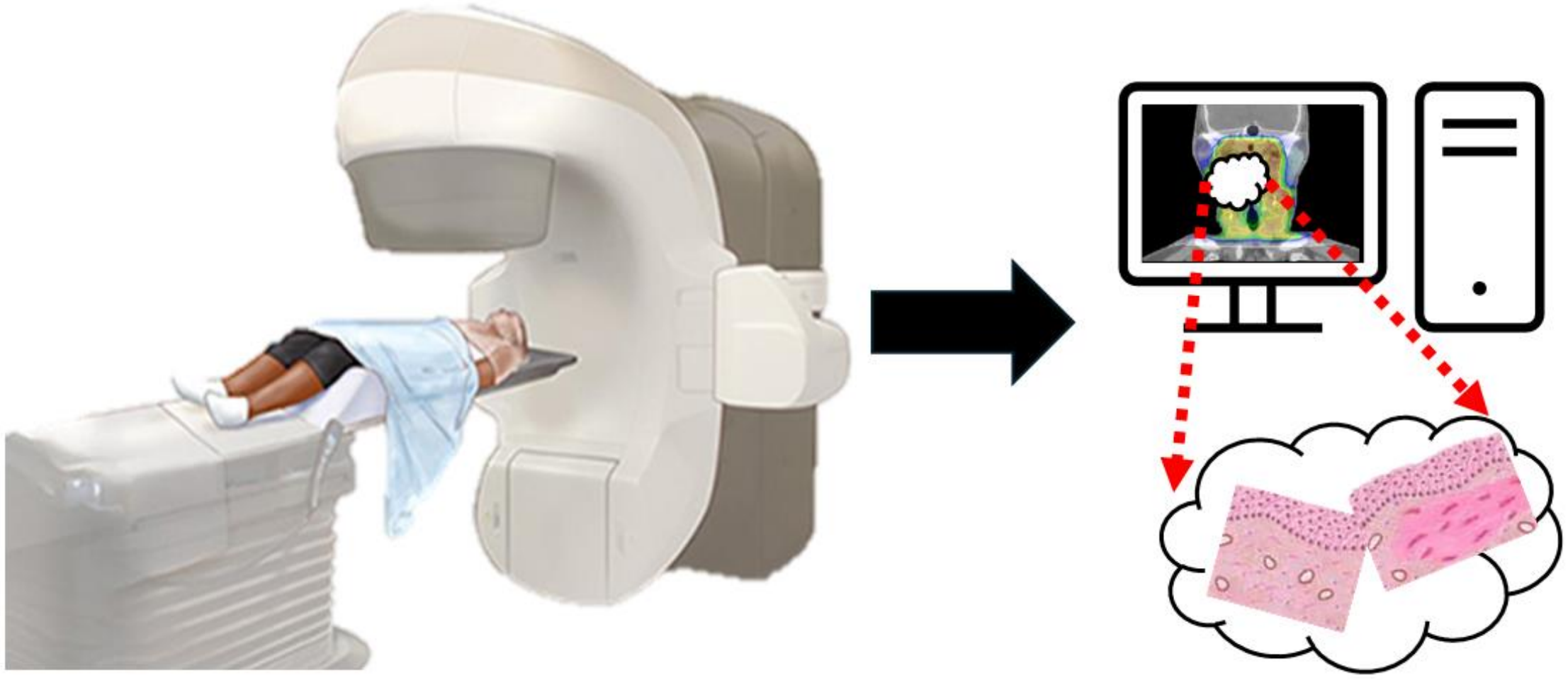

T1ρ as an MRI relaxometric parameter is sensitive to microenvironment and tissue composition changes thus promising to be an effective approach to characterize response to radiation therapy through longitudinal T1ρ characterization of various tissues. T1ρ maps are typically acquired using a spin-lock radiofrequency pulse applied in the direction of the magnetization vector and rotates with the spins at the Larmor frequency[148]. Thus, it represents the longitudinal, T1 relaxation in the rotational frame of reference, which results in the sensitivity to slow moving molecules that are of particular interest in radiation therapy due to the acute structural changes in the extracellular matrix. T1ρ could be leveraged to evaluate tumor response[149] or changes in extracellular matrix in normal tissue reflective of fibrosis development[150], and it has recently been demonstrated to be feasible on radiation therapy devices such as on the 1.5T MR-Linac and to evaluate muscle fibrosis development in response to radiation therapy in rectal patients and could be readily applied to the head and neck region given the considerable dose delivered to functional muscles associated with common toxicities[151]. Additionally, due to the interest in T1ρ for use in the musculoskeletal community, a quantitative assessment profile was published by the Quantitative Imaging Biomarker Alliance (QIBA) to give guidance on acquisition standards using this approach that could be leveraged within the radiation therapy space for effective utilization[152]. One major factor currently limiting effective standardization and implementation is the lack of available reference devices to evaluate quantitative performance in phantoms, like what exists for quantitative measurement of T1 and T2 which has been universally accepted within the MR community[153,154].

### *4.2. MRI-Based Oxygen Measurement: Blood Oxygen Level-Dependent (BOLD) and Oxygen-Enhanced*

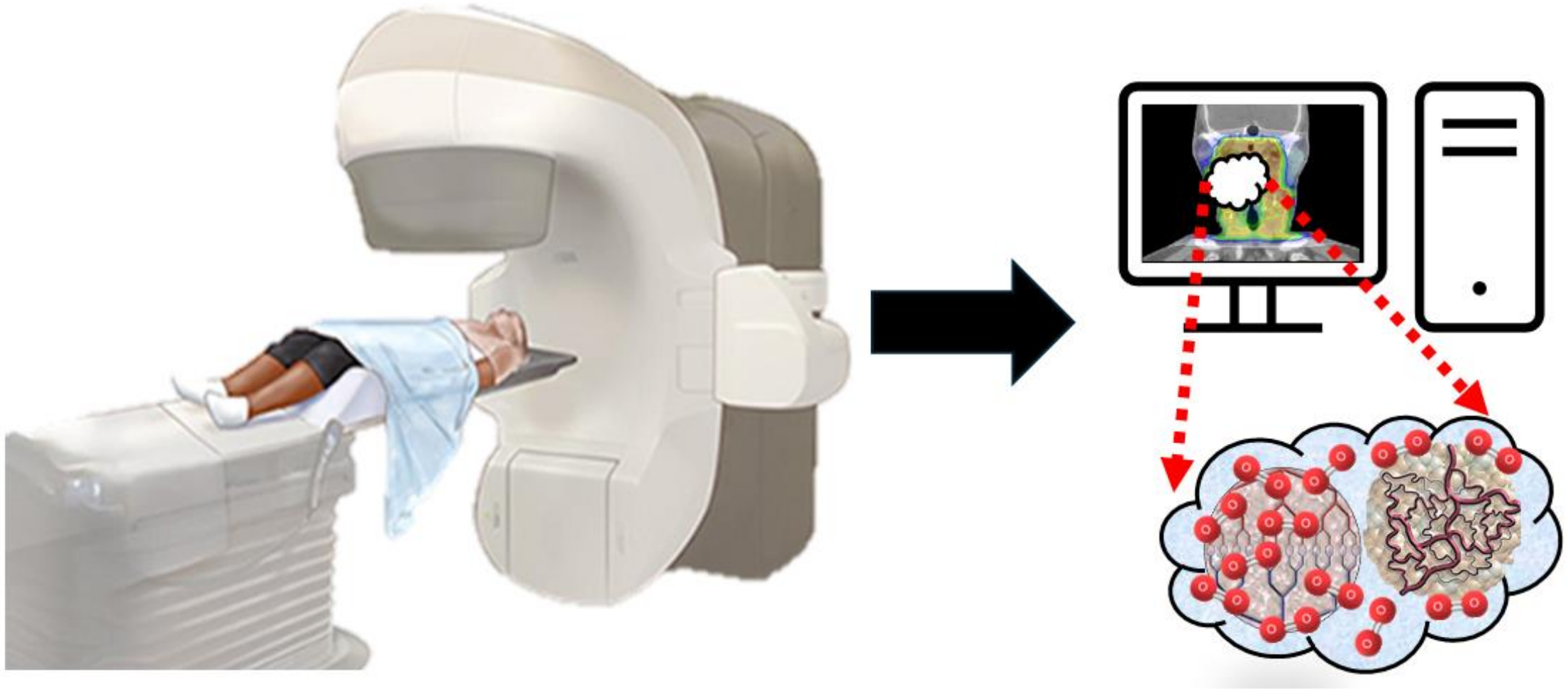

Tumor hypoxia is one of the original hallmarks of cancer[155] and therefore must be studied using a technique which can effectively visualize oxygen content inside a region of interest. One such technique is blood oxygen level-dependent (BOLD) MRI which derives its signal from the marked change in signal intensity from deoxyhemoglobin as the subject externally receives oxygen or carbogen[156]. MR-based biomarkers have been shown to correlate with quantitative measurements of pO2 and can be incorporated in treatment response measurements for ART[157].

In the head and neck, a recent study has shown the feasibility of measuring oxygen changes in the tumor, though the results were inconsistent[158]. Therefore, the repeatability and sensitivity of this method has been studied and found the limit of agreement to be 13% implying a signal intensity change of greater than 10% should be sufficient to detect meaningful differences in oxygenation[159]. This technique has been successfully translated to the MR-Linac in HNC patients, demonstrating significant signal changes between room air and supplemental oxygen[160]. Further, the within-subject coefficient of variation between the diagnostic and MR-Linac machine at 1.5T was comparable, though high, at 25% and 33%, respectively. Another similar study showed promise in deriving the relative oxygen extraction fraction, or rOEF, on the MR-Linac as a potential biomarker for tumor hypoxia[161]. In summary, measuring oxygenation status in tumors and other tissues has shown feasibility in the head and neck, though challenges remain to be addressed before its consistent clinical implementation.

### *4.3. Non-Contrast Perfusion MRI: Arterial Spin Labeling (ASL) and Intravoxel Incoherent Motion (IVIM)*

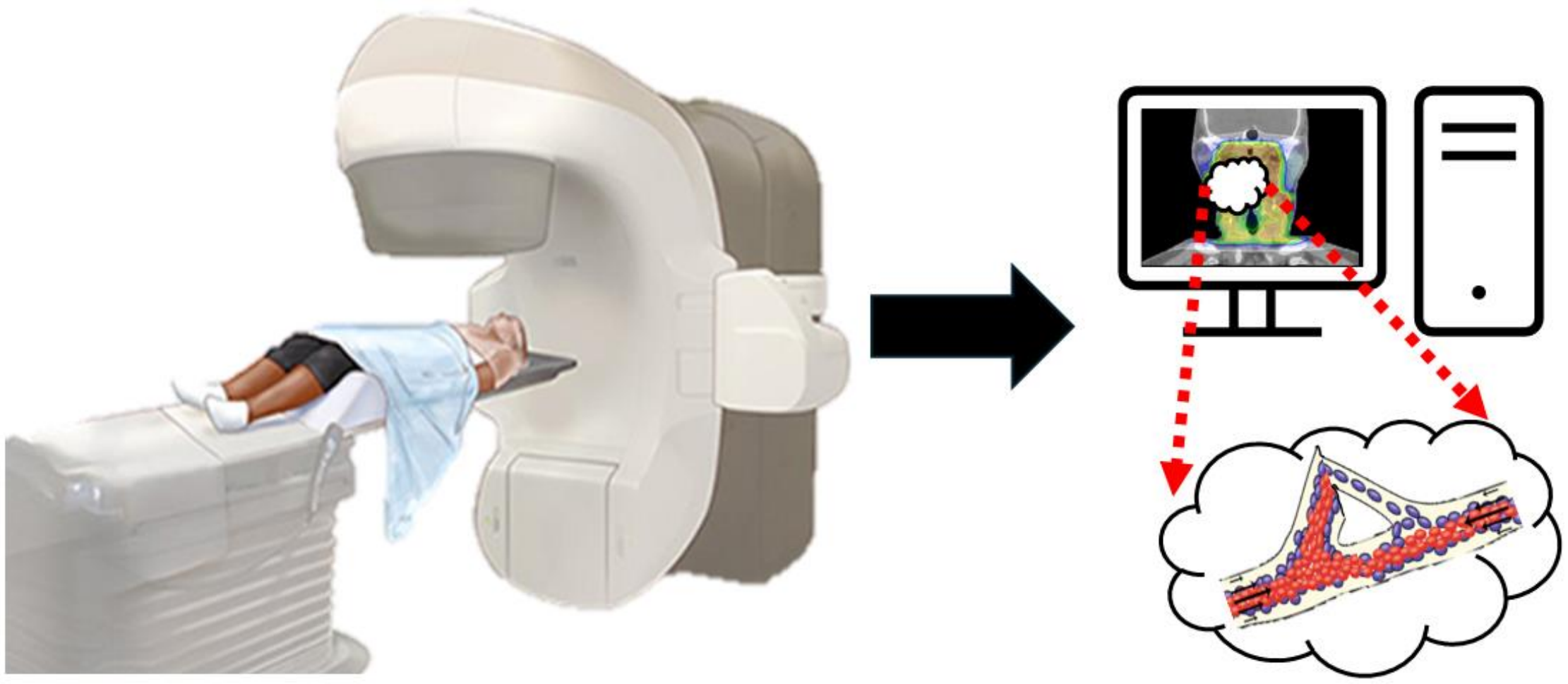

Although DCE MRI is a powerful option for perfusion analysis, its use of an exogenous contrast agent hinders its clinical utility due to the potential harms of gadolinium[162] and the invasive nature of contrast agent injection limiting its validation in phantoms and healthy volunteers. Two primary MRI-based techniques can quantify perfusion without contrast agents: arterial spin labeling (ASL) and intravoxel incoherent motion (IVIM).

ASL, at a fundamental level, is designed using a labeling pulse and a redout scheme[163]. This can be further subdivided into continuous (CASL), pulsed (PASL), and pulsed continuous (PCASL) varieties depending on the application[164]. Some basic tradeoffs are that CASL has high signal-to-noise-ratio (SNR) though suffers from high specific absorption rate (SAR), PASL has lower SNR though its SAR and scan time is reduced, and PCASL combines the best of CASL and PASL with high SNR and low SAR. ASL, in general, has previously been applied in the head and neck to investigate blood flow in salivary glands[165], lesion staging[166,167], and treatment assessment[168]. ASL has recently been investigated on the 1.5T MR-Linac in glioblastoma patients demonstrating general feasibility[169].

IVIM, on the other hand, is a contrast mechanism achieved by acquiring a series of DWI images at varying b-values, specifically focusing on the lower b-values, and fit to an exponentially decaying function of the resulting signal intensities on a voxel level[170]. This function can be defined using several parameters: the perfusion fraction (f), pseudo-diffusion coefficient (D*), and the apparent diffusion coefficient (D). The perfusion fraction is an estimation of the capillary fraction per voxel, the pseudo-diffusion coefficient reflects the disorganized and dense perfusion inside the capillaries, and the apparent diffusion coefficient is the traditional ADC measurement taken from a b-value of 0 s/mm$^2$. In the head and neck, IVIM has been applied to study salivary gland function[171], diagnose and stage tumors[172], assess the levels of perfusion in tumors[173], and as a prognostic test for treatment response[174–176]. On the 1.5T MR-Linac, IVIM has been investigated to assess prostate cancer response to irradiation[177,178], however low correlations were seen due to the large voxel requirement to account for noise making translation to the head and neck difficult.

Diffusion kurtosis imaging, or DKI, could provide another potential biomarker[179] though it is typically achieved using b-values greater than 1,000 s/mm$^2$ which is currently unachievable outside of a research setting with sufficient SNR on the 1.5T MR-Linac (maximum recommendation of 500 s/mm$^2$) in addition to the long scan-time required for sufficient quantification due to the gradient limitations leading to additional noise from motion artifacts[109].

### 4.4. Multiparametric Quantitative Biomarkers

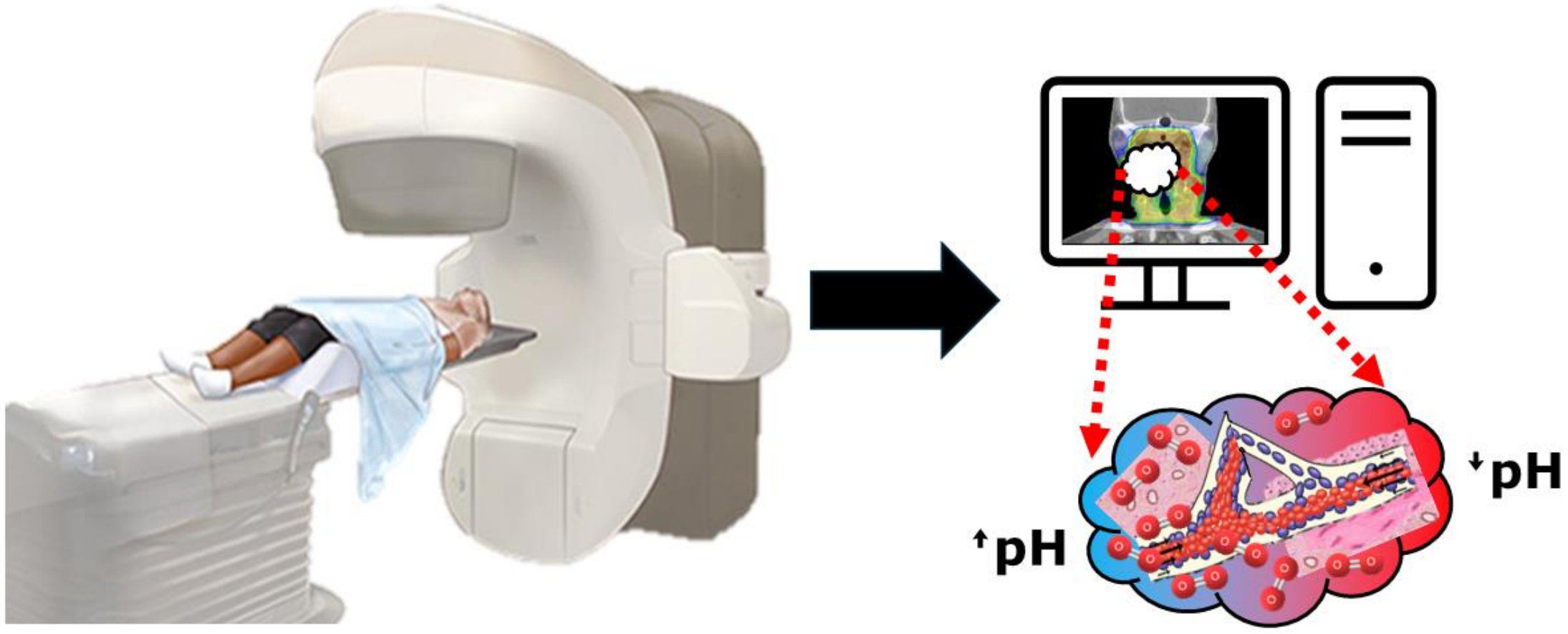

The techniques introduced in this paper can be used individually, though more effectively when combined. This is the concept of multiparametric biomarkers which derive predictive and prognostic conclusions from several sources of information (i.e., combining DWI and DCE), or multiple interpretations from the same source (i.e., radiomics). Both approaches have been utilized in the general HNC space, however the ART area will benefit the most due to its requirement to find solutions to when and how to adapt treatment to maximize tumor control while minimizing normal tissue toxicities.

A recent clinical trial investigated the combination of DWI, DCE, and blood oxygen level-dependent (BOLD) functional information to assess tumor response in mucosal primary HNC as verified by FDG-PET/CT and clinical findings[180]. One derivative study found significant predictive power at an area under the receiver operating characteristic curve (AUC) of 0.83 of the mean change in ADC at week three with local recurrence[181]. The following sub-sections will investigate emerging MRI techniques that combine several MRI-based quantitative measurements using one sequence.

#### 4.4.1. SyntheticMR

One of the most established multiparametric quantitative technique is SyntheticMR (SyntheticMR AB, Linköping, Sweden), a single scan time (i.e., <6 minutes) inherently co-registered acquisition originally known as QRAPMASTER[182] and more regularly known as multi-dynamic multi-echo (MDME) on Siemens MRI scanners, MAGiC on GE scanners, and SyntAc on Philips scanners. Through their postprocessing software, SyMRI, these acquired images can be reconstructed to quantitative T1, T2, and proton density (PD) maps and derivative synthetic contrast maps including: T1-weighted, T2-weighted, PD-weighted, fluid attenuated (FLAIR), phase-sensitive (PSIR), short inversion time (STIR), and double inversion recovery (DIR). Synthetic MR-Sialography images can also be obtained by setting heavy T2-weighting and inversion-based fat suppression which can be applied to study the radiation response in salivary glands to better spare them using adaptive approaches[183].

The most common implementation of the sequence is the 2D-MDME[182,184,185] which acquires multiple 2D slices, often with slice thicknesses of 3 – 6 mm, however, a 3D-isotropic version (3D-QALAS) has been developed[186] and has shown clinically acceptable quantitative accuracy and repeatability in a multi-center[187] and multi-vendor study[188]. Pulse sequence design, technical considerations, and future directions of SyntheticMR can be seen in the review article by Hwang et al. 2022[189].

SyntheticMR has seen increasing usage for diagnostic imaging, however limited investigation has been conducted in usage for radiation oncology[190–193] and even fewer studies have focused on the head and neck[194,195]. Fortunately, the technical feasibility of SyntheticMR in the head and neck radiation oncology workflow has recently been investigated[196–198] indicating its potential for rapid acceleration of MRI-based quantitative biomarkers for ART decision-making.

### *4.4.2. Magnetic Resonance Fingerprinting (MRF)*

Magnetic resonance fingerprinting, or MRF, is a promising innovative approach to combine multiple MRI-based contrasts into one acquisition which can be applied to make adaptive decision[199]. The basic principle to derive T1 and T2 quantitative information is to acquire a dynamic series of time varying flip angles and repetition times (TR), predict their signal evolution behavior using the MRI signal equations (i.e., Bloch equations) across ranges of T1 and T2, encode this information into a dictionary, and match the dictionary to the per-voxel acquired dynamic signal. At the time of this writing, this functionality has been made clinically available on Siemens scanners. Advantages of MRF over combining traditional T1 and T2 mapping are inherent registration between the two maps and a more clinically feasible scan time given a strict accuracy constraint. Similarly, advantages of MRF over the more directly comparable SyntheticMR include robustness of motion and the ability to encode more than just traditional relaxometric data.

For example, recent studies have combined the traditional T1 and T2 encoding with T1-rho with application to the assessment of knee cartilage, however the same approach can be translated to the head and neck for muscle injury assessment[200–202]. Other studies have looked at the combination of CEST with MRF[203,204]. On of the most exciting recent expansions of MRF is the combination of perfusion in terms of arterial spin labeling (ASL), diffusion in terms of ADC, T1, and T2* in a scan time at 75% of the sum of the respective scans individually with the additional inherent co-registration[205]. Combining this information with validated biomarkers like ADC can become the new standard for ART in the head and neck.

MRF in the adaptive radiation oncology space has been limited, though interest has been growing[206]. Several studies have investigated MRF on the low-field 0.35T MR-Linac for general simultaneous T1 and T2 quantification[207] and acceleration of scan times down to three minutes[208]. Only one prior study to the author's knowledge has investigated the clinical feasibility of MRF when combined with the 1.5T MR-Linac for simultaneous T1, T2, and PD quantification[209]. Developing this preliminary study further will accelerate the adoption of multiparametric approaches to help guide ART decisions in the head and neck space. Further, recent innovations of MRF to the 4-dimensional acquisitions could be combined with the recent CMM framework on the 1.5T MR-Linac to provide additional contrasts for real-time tumor and OAR tracking[210].

### *4.5. Normal Tissue Complication Probability (NTCP)-Aware ART*

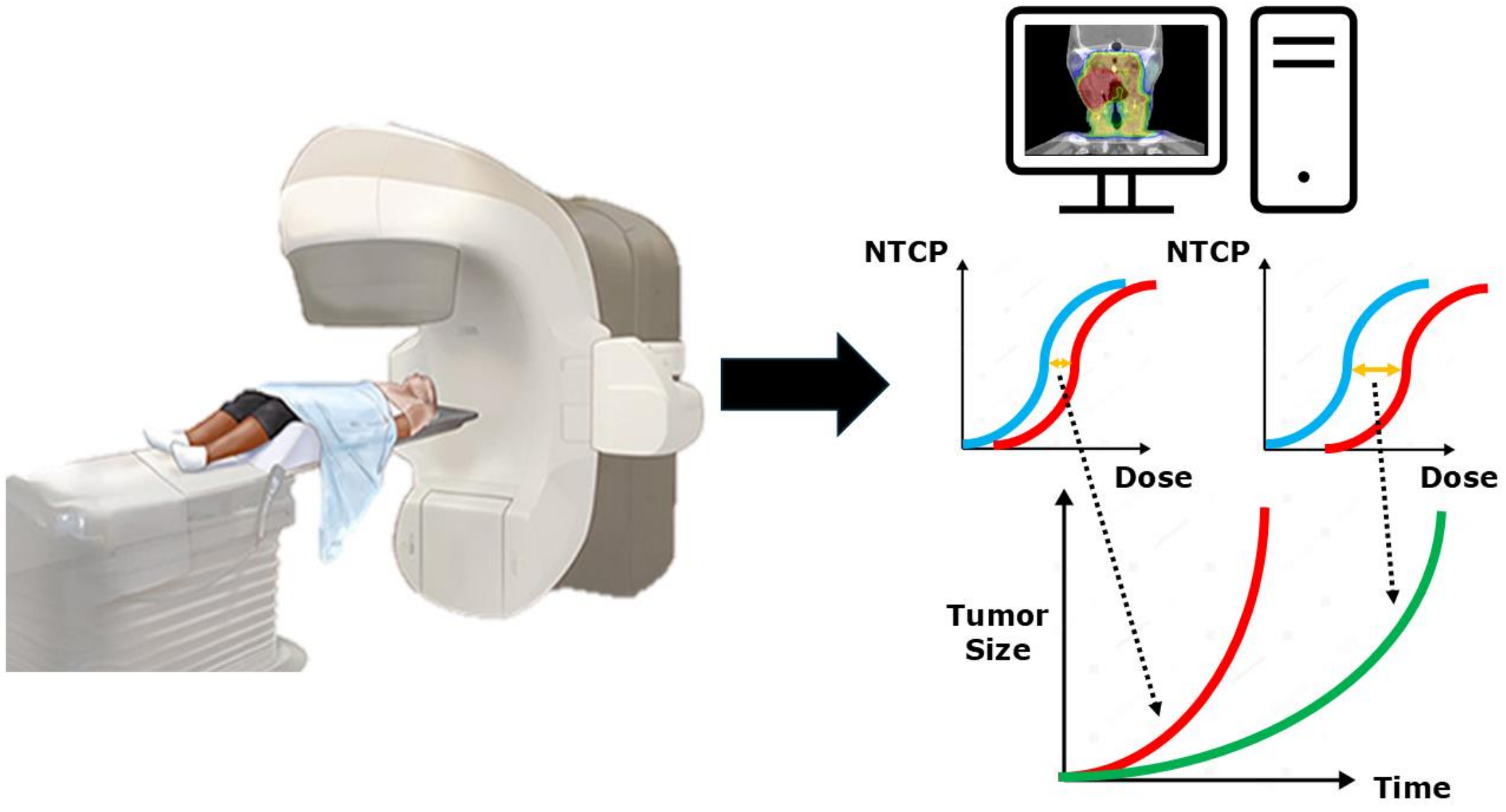

Normal tissue complication probability (NTCP) models predict the likelihood of normal tissue complications resulting from radiation exposure considering factors like the radiation dose, clinical variables, and patient-specific characteristics. By integrating NTCP models into the ART workflow, the treatment can be adapted not only based on anatomical changes but also on predicted toxicity risks to normal tissues. The adaptive process includes feedback from real-time NTCP calculations, where the radiation dose distribution is continuously assessed against the predicted risks of normal tissue complications, allowing for dynamic adjustments. The NTCP of the planned dose distribution is compared to NTCP from the actual dose distribution, as calculated from the delivered treatment fraction and the anatomical information obtained from the online image acquisition for that fraction. If the actual NTCP exceeds a certain threshold or deviates significantly from the planned NTCP, adaptive adjustments can be made. However, determining this delta-NTCP threshold can be challenging. A retrospective study by Heukelom et al. 2020 on 52 patients with HNC showed that the use of delta-NTCP as an objective selection strategy for ART is superior to clinical judgment[211]. A study by Gan et al. 2024 suggested the 3rd week of treatment as the generic optimal re-planning time for a set of 10 HNC OARs to ensure a limited increase of accumulated mean dose of 3 Gy simultaneously for multiple organs[212]. As acknowledged by the authors, NTCP-based thresholding is preferable as the same dose-base threshold can result in different toxicity risks. A study by Nosrat et al. 2024 aimed to determine the optimal threshold over the treatment course by calculating the delta-NTCP at different time points (at 10, 15, 20 and 25 delivered fractions) for multiple HNC toxicities (xerostomia, dysphagia, parotid gland dysfunction, and feeding tube dependency at 6 months post-RT)[213]. Based on the calculated delta-NTCP values, the optimal time for re-planning was determined using a Markov decision process (MDP) model.

All these studies are based on CT imaging, however NTCP-aware MRI-guided ART (MRIgART) is a less commonly studied field. MRI is a non-ionizing imaging modality, allowing for repeated scans without additional radiation exposure. However, while MRI also offers better visualization of normal tissues compared to x-ray-based imaging modalities, MRI pixel values do not contain the quantitative information (electron density) used for dose calculation[73]. This additional challenge of MRIgART coupled with the less widely available MRI imaging resources (compared to CT), have contributed to the slower adoption of NTCP aware MRIgART.

Looking ahead, the integration of NTCP models into MRIgART holds significant promise for further enhancing the precision of cancer treatment. Advances in synthetic CT generation from MRI data, combined with improved algorithms for MRI-based dose calculations, are likely to overcome current limitations, enabling more widespread adoption of NTCP-aware MRIgART.

### *4.6. Optimization-Based Frameworks*

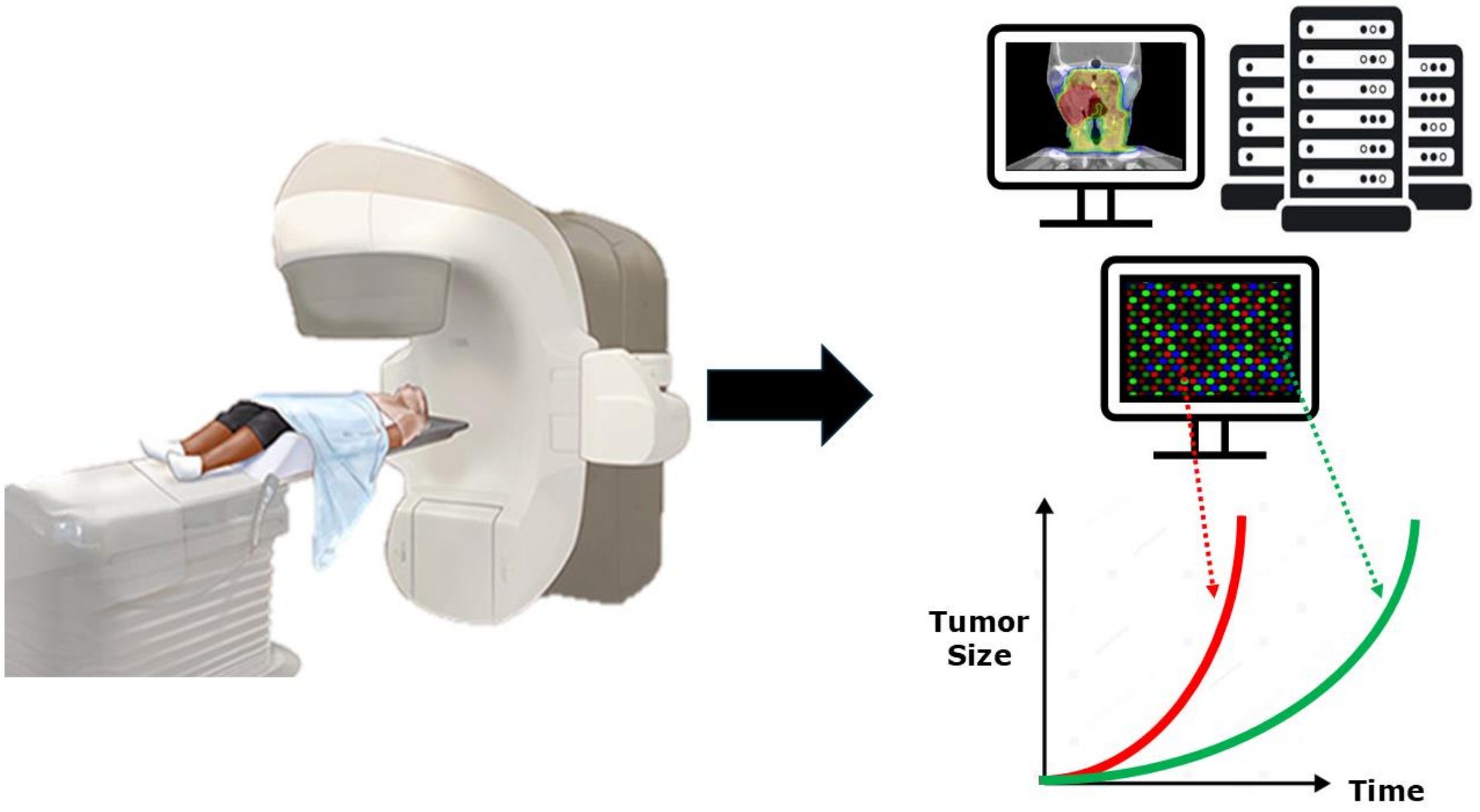

Clinical decision making in the ART setting is infinitely complex. Wide decision spaces are available due to the high number of fractionations, continuous dose levels, and vast resources for biomarker-based adaptations. Markov models, a potential solution to this problem, are mathematical optimization frameworks for stochastic systems specialized for finding the optimal policy to vast decision spaces, such as in ART.

These models have recently been proposed to optimize the ART workflow by incorporating strict rules on transitions of patients from one state to another[214]. Similar mathematical models have been applied to dose-volume histograms (DVH) using an unsupervised clustering algorithm to identify risk levels for osteoradionecrosis[215]. The Markov decision process, or MDP, has the potential to guide optimal re-planning scenarios given a constrained problem-space and unlimited solution-space which is becoming more prevalent as MRI-biomarker techniques advance[213]. This approach has already been tested using a dataset of over 1,500 HNC patients treated over a 10-year period to determine the optimal times to scan patients for surveillance following radiation therapy[216,217].

### *4.7. Digital Twins*

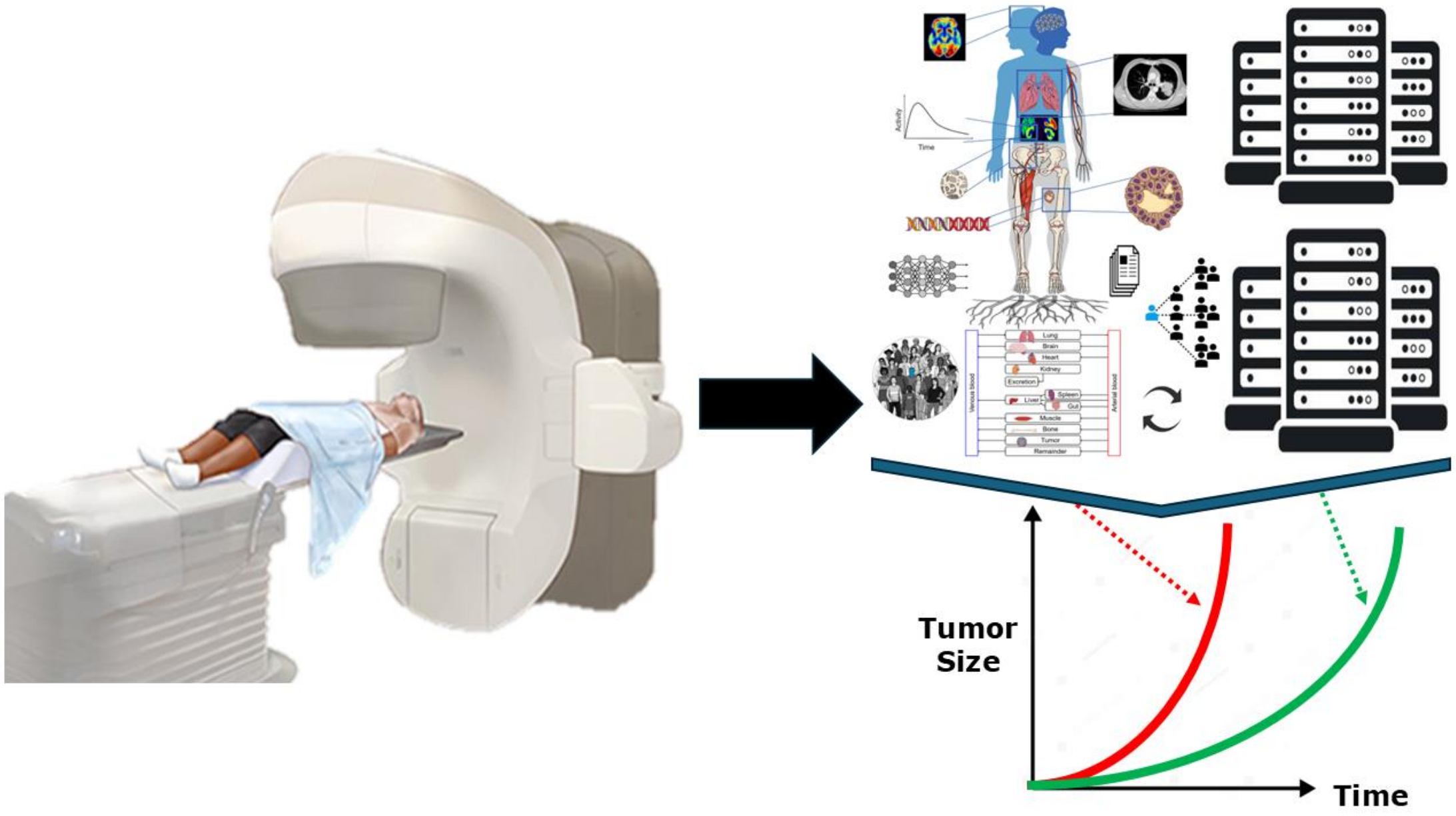

Digital twin technology, initially conceptualized in the 1960s and significantly advanced in the early 21st century, has evolved into a powerful tool for creating virtual replicas of physical objects or systems for real-time analysis, simulation, and optimization. The Digital Twin Consortium defines it as a virtual representation of a real-world object or system that is continuously updated to reflect real-time behaviors and states[218].

In radiation oncology, digital twins offer the potential to create highly personalized treatment plans by integrating multimodal, patient-specific data, aligning with the principles of precision medicine[219]. This approach aims to enhance treatment outcomes by minimizing damage to healthy tissues while maximizing the efficacy of targeting cancer cells. The application of digital twins in radiation oncology involves creating virtual patient models, real-time analysis and simulation, and predictive modeling leveraging biophysical modeling and machine learning. Digital twin technology and ART offer solutions to the current challenges in radiation oncology by addressing anatomical changes, improving dose delivery, personalizing treatment plans, and enabling dose escalation[220,221]. The same studies have shown the dynamic changes in tumor size and shape over radiation for HNSCC, highlighting the need for personalized adaptive approaches.

The integration of digital twins with other advanced technologies has further enhanced their capabilities in radiation oncology. IGRT combined with digital twins has shown significant improvements in the accuracy and efficacy of radiation dose delivery[222]. AI algorithms are being used for auto-contouring, dose calculations, and treatment plan optimization[220,223]. Internet of Things (IoT) technologies are being incorporated to provide real-time data acquisition, allowing for dynamic and continuous updates to digital twin models[218]. Despite its promising benefits, implementing digital twin technology in HNC ART presents several challenges. These include data integration and standardization, model complexity and interpretability, uncertainty handling, data requirements and privacy, and interdisciplinary collaboration[218]. Recent advancements aim to address these challenges through data assimilation methods, real-time data acquisition, and leveraging AI and machine learning.

The future of digital twin technology in radiation oncology is promising, with potential for enhanced predictive accuracy, operational optimization, and convergence with other technologies[224]. Ongoing research aims and collaborative efforts, such as the NCI-DOE Collaboration and initiatives like the Cancer Moonshot, are crucial for fostering interdisciplinary partnerships and driving advancements in this field[218]. While the potential benefits of digital twin technology in radiation oncology are significant, several ethical considerations and challenges need to be addressed. These include data privacy and security, equitable access to advanced healthcare technologies, interpretability of AI models, validation and regulatory approval, and training and education for healthcare professionals.

## 5. Conclusion and Reflections

In conclusion, the future of ART in head and neck cancer is just beginning. Novel technologies have pushed the boundary of what is possible in terms of techniques to identify biomarkers for adaptation as well as innovative devices specialized to respond to these adaptations, sometimes in real-time. Important interdisciplinary steps must be taken moving forward to ensure the safe deployment of these new techniques, such as rigorous quality assurance evaluations from medical physicists, clinical trials from physicians, and comprehensive testing from vendors prior to release. In summary, we aimed not to provide a single correct answer for the optimal implementation of ART in the era of imaging biomarkers, but to encourage the field to collaborate and bring each idea discussed here together to overcome current barriers and deliver the best treatment possible to the patient.

The development and validation of MR-based imaging biomarkers currently suffers from a lack of standardized methodology and consensus, leading to inconsistencies in how these tools are evaluated and reported. Technologies such as DCE-MRI and MR fingerprinting are difficult to place within existing evaluation frameworks (e.g., clinical trials, R-IDEAL[32], or DECIDE-AI[31]), as they could be considered biomarkers, AI-driven software, or medical devices. This gap underscores the pressing need for a structured conceptual framework, to standardize methods and nomenclature within the context of imaging biomarkers in radiation oncology applications. Moreover, as organizations like RSNA shift their focus, including a move away from QIBA initiatives, a clear gap has emerged in the development and leadership of imaging biomarkers relevant to radiation oncology. With imaging biomarkers playing an increasingly critical role in precision radiotherapy, it is no longer appropriate to rely solely on the diagnostic imaging community to drive their advancement. A Delphi study could serve as a starting point to build consensus and lay the foundation for an "IB-IDEAL" initiative for a harmonized development pathway for imaging biomarkers in radiation oncology.